\documentclass[onecolumn,aps,nofootinbib,superscriptaddress,tightenlines,notitlepage,pra]{revtex4-2}
\usepackage[english]{babel}
\usepackage{amsmath} 				
\usepackage{amssymb} 				
\usepackage{amsthm}
\usepackage{array}       				
\usepackage{graphicx}  	
\usepackage{colortbl}
\usepackage{xcolor}
\usepackage[utf8]{inputenc} 			
\usepackage{url}     
\usepackage{times}
\usepackage{caption}
\usepackage[shortlabels]{enumitem}
\usepackage{braket}
\usepackage{mathtools}
\usepackage{pifont}
\usepackage{verbatim}
\usepackage{tikz}
\usetikzlibrary{positioning}
\usepackage{subcaption}

\usepackage{hyperref}
\hypersetup{
    colorlinks=true
}
\begin{document}
\makeatletter

    \def\CT@@do@color{%
      \global\let\CT@do@color\relax
            \@tempdima\wd\z@
            \advance\@tempdima\@tempdimb
            \advance\@tempdima\@tempdimc
    \advance\@tempdimb\tabcolsep
    \advance\@tempdimc\tabcolsep
    \advance\@tempdima2\tabcolsep
            \kern-\@tempdimb
            \leaders\vrule
                    \hskip\@tempdima\@plus  1fill
            \kern-\@tempdimc
            \hskip-\wd\z@ \@plus -1fill }
    \makeatother
\newcommand{\id}[0]{\mathrm{I}}
\newcommand{\simulator}[0]{\textbf{pylove} }
\theoremstyle{plain}
\newtheorem{clm}{Claim}
\newcommand{\codeline}[1]{{\color{black!34} $\mathtt{#1}$}}
\title{Simulating quantum error mitigation in fermionic encodings}
\author{R. W. Chien}
\affiliation{Department of Physics and Astronomy, Dartmouth College, Hanover, NH 03755, USA}

\author{K. Setia}
\affiliation{qBraid, 5235, Harper Court, Chicago, IL, USA}
\affiliation{Department of Physics and Astronomy, Dartmouth College, Hanover, NH 03755, USA}

\author{X. Bonet-Monroig}
\affiliation{Instituut-Lorentz, Universiteit Leiden, P.O. Box 9506, 2300 RA Leiden, The Netherlands}

\author{M. Steudtner}
\affiliation{Dahlem Center for Complex Quantum Systems, Freie Universität Berlin, 14195 Berlin, Germany}
\affiliation{Instituut-Lorentz, Universiteit Leiden, P.O. Box 9506, 2300 RA Leiden, The Netherlands}
\affiliation{QuTech, Delft University of Technology, Lorentzweg 1, 2628 CJ Delft, The Netherlands}

\author{J. D. Whitfield}
\affiliation{Department of Physics and Astronomy, Dartmouth College, Hanover, NH 03755, USA}
\date{\today}

\begin{abstract}
    The most scalable proposed methods of simulating lattice fermions on noisy quantum computers employ encodings that eliminate nonlocal operators using a constant factor more qubits and a nontrivial stabilizer group. In this work, we investigated the most straightforward error mitigation strategy using the stabilizer group, stabilizer postselection, that is very natural to the setting of fermionic quantum simulation. We numerically investigate the performance of the error mitigation strategy on a range of systems containing up to 42 qubits and on a number of fundamental quantum simulation tasks including non-equilibrium dynamics and variational ground state calculations. We find that at reasonable noise rates and system sizes, the fidelity of computations can be increased significantly beyond what can be achieved with the standard Jordan-Wigner transformation at the cost of increasing the number of shots by less than a factor of 10, potentially providing a meaningful boost to near-term quantum simulations. Our simulations are enabled by new classical simulation algorithms that scale with the logical Hilbert space dimension rather than the physical Hilbert space dimension.
\end{abstract}

\maketitle

\section{Introduction}
\begin{table}[t]
    \centering
    \begin{tabular}{lrcccl}
\textbf{Code family (Abbreviation)}& \textbf{Ref}&  $\,\,\,$\textbf{d=2}$\,\,\,$ & \textbf{d=3}  &$\;$& \textbf{Original authors}  \vspace{.1cm}\\
         \rowcolor{black!8}Original superfast encoding (\textbf{BKSF})& \cite{bravyi2002fermionic} &  {\color{black!40}\ding{51}} &   && Bravyi \& Kitaev\\
         Auxiliary fermion code & \cite{ball2005fermions, verstraete2005mapping,whitfield2016local} &  {\color{black!40}\ding{55}} & {\color{black!40}\ding{55}}  && Ball / Verstraete \& Cirac\\
         \rowcolor{black!8} Exact bosonization (\textbf{EB})& \cite{chen2018exact} &  {\color{black!40}\ding{51}} &  && Chen, et. al\\
         Auxiliary qubit code (\textbf{AQM})& \cite{steudtner2019quantum,chiew2021optimal,o2022ultrafast} & {\color{black!40}\ding{55}}& {\color{black!40}\ding{55}} && Steudtner \& Wehner \\
         \rowcolor{black!8}Generalized superfast encoding (\textbf{GSE}) & \cite{setia2019superfast} & {\color{black!40}\ding{51}} &{\color{black!40}\ding{51}}  && Setia, Bravyi et al{.} \\
        Majorana loop stabilizer code & \cite{jiang2019majorana} & {\color{black!40}\ding{51}}& {\color{black!40}\ding{51}} && Jiang, McClean et al{.}\\
         \rowcolor{black!8}Compact encodings (\textbf{Compact}) & \cite{derby2021compact,bausch2020mitigating,derby2021compact2}&  {\color{black!40}\ding{55}} & {\color{black!40}\ding{55}} && Derby \& Klassen \\
         Tetron/Hexon code & \cite{landahl2021logical} & {\color{black!40}\ding{51}} & {\color{black!40}\ding{51}} && Landahl \& Morrison \\
         \rowcolor{black!8} Super-compact code &\cite{chen2022equivalence}& {\color{black!40}\ding{55}} & {\color{black!40}\ding{55}} && Chen \& Xu \\

 Optimized encodings &  \cite{chien2022optimizing} & {\color{black!40}\ding{51}} & {\color{black!40}\ding{55}} && Chien \& Klassen\\
\rowcolor{black!8} Higher distance bosonization & \cite{chen2022error}  &   {\color{black!40}\ding{51}} & {\color{black!40}\ding{51}} && Chen, Groshkov \& Xu \\
         Generalized auxiliary qubit code (\textbf{GAQM}) & \hyperref[sec:genaux]{[here]}  & {\color{black!40}\ding{51}} & $\,$  && $\,$ 
    \end{tabular}
    \caption{Zoo of quantum codes for fermion-to-qubit mappings and their error-mitigation properties. We showcase the error mitigation properties of each code.  Where known, we indicate whether  scalable instances of each code exist that are at least distance two (three), such that they would allow us to detect (correct) single qubit errors. Note that some of these codes are not meant to detect errors, but were designed to preserve operator locality in the fermion-to-qubit mapping. A description of the Generalized auxiliary qubit code can be found in Appendix \ref{sec:genaux}.}
    \label{tab:codes}
\end{table}
Quantum computers are heralded to afford us the ability to study quantum systems in ways beyond the capabilities of classical computers. This is particularly the case for the simulation of strongly interacting fermions, giving the potential to advance the fields of condensed matter physics, particle physics, and quantum chemistry. However, many of the most promising quantum simulation algorithms will require executing circuits well beyond the capabilities of today's intermediate scale, noisy quantum devices. Computational errors stemming from noise effectively limit the depth of quantum algorithms that can be run such that their computational outcome can still be learned.
While quantum error correction offers the ability to suppress such errors, and progress towards its experimental realization has been made \cite{hu2019quantum,andersen2020repeated,egan2021fault,QuantinuumEC,google2023suppressing}, fault-tolerant quantum computation remains at this point out of reach. This is due to the fact that quantum error correction requires a large number of physical qubits and low noise rates to form even one logical (computational) qubit. 

Without access to quantum error correction, it is unreasonable to expect to be able to perform arbitrarily long computations. More pragmatically, we may seek to mitigate the effects of errors. Indeed, the field of quantum error mitigation seeks ways to judiciously spend resources (such as additional samples) to extract useful signal from a noisy computation. While this might incur an overhead in the total number of computations to be run, it does not come with the stringent hardware requirements of quantum error correction. Early error mitigation schemes that continue to see extensive use first arose in the field of quantum control \cite{viola1999dynamical} and recently proposed schemes  include data-driven \cite{temme2017error,lowe2021unified}, distillation-based \cite{huggins2021virtual}, and symmetry-based approaches \cite{bonet2018low,mcclean2020decoding}.
 
Symmetry-based error mitigation is particularly interesting, as it seeks to filter the results of corrupted experiments identified by observing the computational state to be in an incorrect symmetry sector. These symmetries can either occur naturally \cite{mcardle2019error, bonet2018low}, or can arise as stabilizer conditions of quantum codes \cite{setia2019superfast,derby2021compact} as in quantum error correction.

The viability of these codes is however hard to test with classical computers, particularly if we want to investigate their error mitigation ability in circuits resembling actual quantum computations. We are usually interested in algorithms that are explicitly non-Clifford, and the qubit overhead, albeit modest, is usually enough to hamper state-vector simulators. 

In this work, we classically simulate error mitigation schemes based on symmetry verification of locality-preserving fermionic encodings. Using a novel simulator for systems with stabilizer symmetries, we test the performance of various quantum codes within noisy quantum simulation experiments.

These experiments typically take the form of arbitrary-angle rotations generated by logical operators -- Pauli strings acting on logical qubits -- implemented by rotations of the equivalent Pauli string on physical qubits. A system with $n$ physical qubits and $k$ generators of the stabilizer group has $n-k$ logical qubits. Without noise, it would be easy to model this $n$-qubit physical system with only $n-k$-qubits. That means a state vector simulator would be able to simulate the system in time $O(\mathrm{poly}(2^{n-k}))$ rather than $O(\mathrm{poly}(2^{n}))$. We show that, under a modest overhead, it is possible to reconstruct the simulation of the same system under Pauli noise, as if it only had $n-k$ qubits.

\begin{clm}
\label{clm:simulator}
Acting on a system of $n$ qubits with $n-k$ stabilizer generators, a circuit of $\ell$ rotations generated by logical Pauli operators can be simulated in runtime $\mathrm{poly}(n,n-k,\ell,2^{k})$ per shot.

\end{clm}

On the basis of this work, we have developed the \simulator  (\textbf{py}thon-based \textbf{lo}gical \textbf{ve}ctor) simulator available on github.\footnote{\texttt{https://github.com/msteudtner/pylove\_simulator}}
While the simulator would work with all kinds of stabilizer codes, we want to study the error mitigation properties of fermionic quantum codes in particular. This is because fermionic quantum codes generally have a low distance and no transversal gates. The primary use of these quantum codes is to map fermionic model Hamiltonians to Hamiltonians acting on qubits in a way that conserves the locality of the problem. Fermionic quantum codes allow for fermionic operators acting on a constant number of geometrically-local fermionic modes to be mapped to operators acting on a constant number of geometrically-local qubits  \cite{bravyi2002fermionic} (for two- or higher dimensional models \cite{chen2020arbitrary,derby2021compact2}). Constant-depth logical fermionic circuits can therefore be simulated with constant-depth qubit circuits. This property is beneficial for reducing the effects of errors because logical operations made of shorter circuits acting on fewer qubits mean fewer opportunities for errors to accumulate during the computation. However, constructing logical operators with small weight is seemingly at odds with coding theory, which tell us that protection against errors is achieved through large code distances (a high minimum Pauli weight of all logical operators). While non-locality does not automatically mean a higher code distance, there is certainly a trade-off between the error mitigation properties and operator weights. 

\begin{clm}
\label{clm:results}
Stabilizer postselection error mitigation combined with locality-preserving fermionic encodings can increase the simulation fidelity of two-dimensional local fermionic circuits beyond Jordan-Wigner at near-term simulation sizes and noise rates with less than an order of magnitude increase in the number of required samples.
\end{clm}

To date, a number of fermionic quantum codes are known, and an overview can be found in Table \ref{tab:codes}. We numerically investigate stabilizer postselection error mitigation in a number of them in the settings of random quantum circuits of fermionic gates, variational ground state preparation with variational quantum eigensolver (VQE), and simulation of a simple finely-tuned model of dynamics originally proposed in the study of anomalous edge modes in Floquet condensed matter systems.

The organization of the paper is as follows. In Section (\ref{encoding_overview}), we provide a review of fermionic encodings highlighting the stabilizer code structure that will be crucial later. In Section (\ref{techniques}), we introduce the tapering techniques used in the simulator before discussing our simulation algorithms in Section (\ref{simulator}). In Section (\ref{numerical_experiments}), we discuss the numerical experiments we performed to explore the performance of stabilizer postselection error mitigation using fermionic encodings.

\section{Fermion-to-qubit mappings as quantum codes} \label{encoding_overview}
Before we consider the error mitigation abilities of quantum codes for fermionic encodings, we will discuss their use as fermion-to-qubit mappings.  Relevant fermionic operators in molecular and condensed matter systems often take the form
\begin{figure}
    \centering
    \includegraphics[scale=.6]{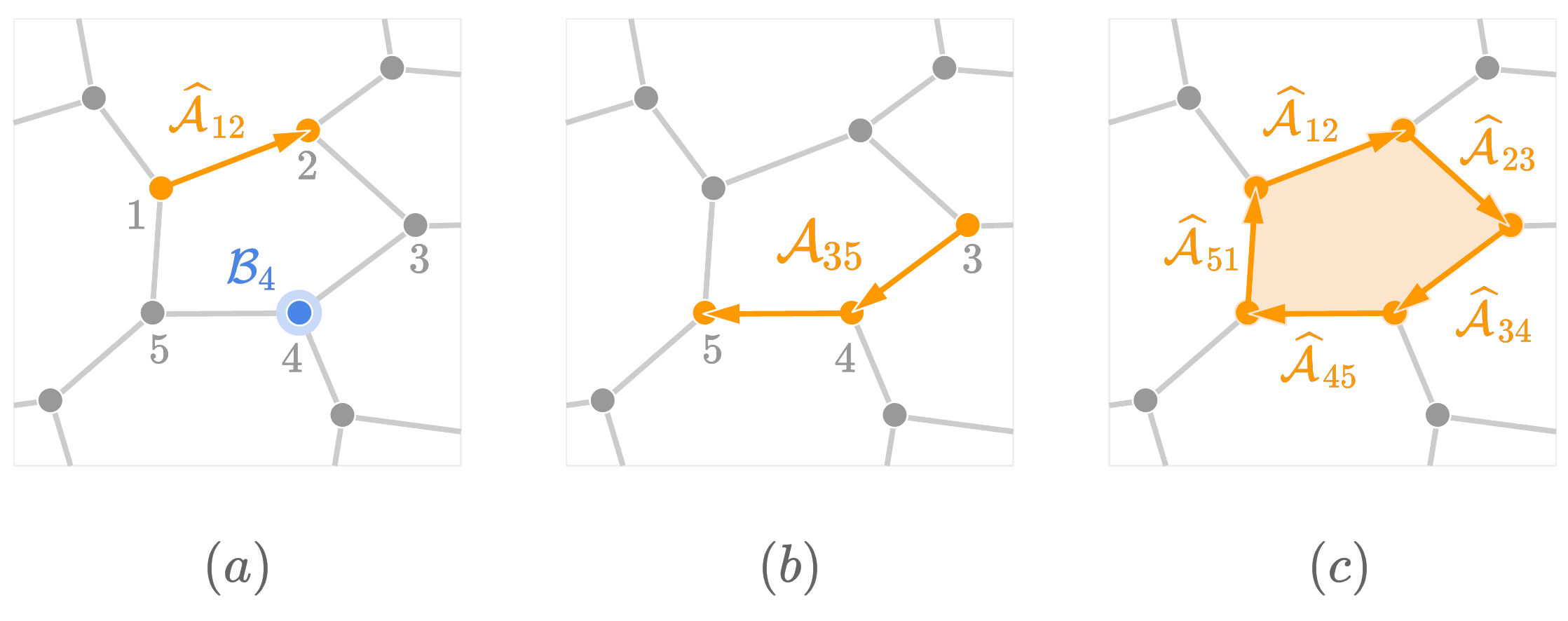}
    \caption{Quantum codes for fermion-to-qubit mappings. $(a)$ Every code is defined on a graph in which every vertex is a fermionic mode. The quantum code itself is a set of vertex operators $\mathcal{B}_k$ (blue) and edge operators $\widehat{\mathcal{A}}_{jk}$ (orange), a generating subset of the exchange operators ${\mathcal{A}}_{jk}$. $(b)$ Exchange operators are defined between any two vertices of the graph, not just vertices connected by edges. However, exchange operators are generally defined by the multiplication of adjacent edge operators. Here we have $\mathcal{A}_{35} = i \widehat{\mathcal{A}}_{34}\cdot\widehat{\mathcal{A}}_{45}$ (orange). $(c)$ Stabilizers are obtained by edge operator products around closed loops. The stabilizer here is $i\mathcal{A}_{11} = i\widehat{\mathcal{A}}_{12}\widehat{\mathcal{A}}_{23}\widehat{\mathcal{A}}_{34}\widehat{\mathcal{A}}_{45}\widehat{\mathcal{A}}_{51}$ (orange). }
    \label{fig:codes}
\end{figure}
\begin{align}
\label{eq:fermop}
    \sum_{j,k} h_{jk}\; a^{\dagger}_ja^{\phantom{\dagger}}_k\; + \; \sum_{j,k,l,m} h_{j,k,l,m}\; a^{\dagger}_ja^{\phantom{\dagger}}_k a^{\dagger}_la^{\phantom{\dagger}}_m\, ,
\end{align}
where $h_{jk}$ and $h_{jklm}$ are arbitrary coefficients and $a^{\dagger}_j$, $a^{\phantom{\dagger}}_k$ are fermion creation and annihilation operators on modes $j$ and $k$, respectively. Every creation-annihilation operator pair can be mapped to a linear combination of signed Pauli strings: each code in Table \ref{tab:codes} provides logical operators
$\mathcal{B}_k$, $\mathcal{A}_{jk}$, called vertex and edge operators\footnote{These operators have also been denoted $V_k$ for vertex operators and $E_{jk}$ for edge operators.}. Vertex and edge operators are signed Pauli strings, and are contained in the set of logical operators of a quantum code underlying the simulation. The quantum code is a fermion-to-qubit mapping, or fermionic encoding, if its vertex and edge operators describe a mapping
\begin{align}
\label{eq:map1}
    a^\dagger_j a^{\phantom{\dagger}}_k \quad \mapsto \quad  \frac{i}{4} (1-\mathcal{B}_j) \mathcal{A}_{jk}(1-\mathcal{B}_k) \, ,
\end{align}
 that can be applied to all pairs of operators in \eqref{eq:fermop}, preserving the fermionic exchange statistics. For this to be the case, vertex and edge operators of  any three distinct indices $(j,k,l): j \neq k \neq l$ and $j\neq l$, vertex and edge operators must obey the following relations:
\begin{align}
\mathcal{A}_{jk} & =  - \mathcal{A}_{kj} \, , \\
\mathcal{A}_{jk} \mathcal{B}_k & =  - \mathcal{B}_k \mathcal{A}_{jk} \, , \\
\mathcal{A}_{jk} \mathcal{A}_{kl}  & =  - \mathcal{A}_{kl} \mathcal{A}_{jk} \, .
\end{align}
To apply \eqref{eq:map1} to number operators, we allow for the case of $j=k$ in which the edge operator $\mathcal{A}_{kk}$ is trivial. We will see very soon that this would yield  $\mathcal{A}_{kk} = -iS$, where $S$ is a stabilizer that would commute with all logical operators. Without loss of generality  we can set $\mathcal{A}_{kk}=-i\,\id$ for all $k$ such that
\begin{align} 
\label{eq:mapping}
    a^\dagger_k a^{\phantom{\dagger}}_k \quad \mapsto \quad  \frac{1}{2}(1-\mathcal{B}_k) \, .
\end{align}
Let now us reflect on generating sets of edge operators, and how they can generate the code stabilizers. Vertex and exchange operators are defined on a connectivity graph between the fermionic modes, set by the underlying quantum code. The graph is a set of vertices and edges $(V,E)$, where $V \in \left\lbrace 1, \,\dots \, , N\right\rbrace$ labels the modes, and $(j,k) \in E$ are  connections between two of them. The quantum code then defines a vertex operator ${B}_k$ for every $k\in V$, and a set of exchange operators along the edges. For every edge $(j,k) \in E$ the quantum code provides us with an edge operator $ \widehat{\mathcal{A}}_{jk}$, see Figure \ref{fig:codes}$(a)$. The edge operators $\widehat{\mathcal{A}}$ are the generating set of the exchange operators $\mathcal{A}$: operators between non-adjacent vertices are defined by a product of several $\widehat{\mathcal{A}}$ operators, see Figure \ref{fig:codes}$(b)$. A nonlocal operator $\mathcal{A}_{j_1 j_\ell}$ is proportional to the product of operators along any length-$\ell$ chain of connected edges $\bigcup_{m=1}^{\ell-1} (j_m, j_{m+1}) \subset E $:
\begin{align}
    \label{eq:chain}
    \mathcal{A}_{j_1 j_\ell}  = i^{\ell-1}\prod_{m=1}^{\ell-1}  \widehat{\mathcal{A}}_{j_m j_{m+1}}  \, .
\end{align}
Finally, the quantum code's stabilizers $S$ are can be found by  $ S =  i \mathcal{A}_{j_1 j_\ell}$ for all exchange operators \eqref{eq:chain} with edges forming closed loops such that $j_1 = j_\ell$, see Figure \ref{fig:codes}$(c)$. The set of stabilizers is generated by all sequences $j_1, \, ... ,\, j_{\ell}$ of arbitrary length $\ell$. There is an irreducible set of loops that make up the set of stabilizer generators. 

Note that all the statements in this section also hold for fermion-to-qubit mappings that are not quantum codes. These mappings define edge and vertex operators on a graph without closed loops, such that there are no stabilizers. Infamously, there is the Jordan-Wigner transform \cite{wigner1928paulische} (abbreviated \textbf{JW} in this manuscript), whose connectivity graph is a one-dimensional chain with $\widehat{\mathcal{A}}_{k, k+1} = Y_kX_{k+1}$ and $\mathcal{B}_k=Z_k$ for all $k\in V$.

\section{Techniques} \label{techniques}
The techniques in this paper rely on our ability to remove qubits and stabilizer conditions from the system. 
Stabilizers determine the subspace of the physical Hilbert space which can be used for computation on the logical level. In a system with many stabilizer conditions, the logical subspace is an exponentially small fraction of the full Hilbert space. Hence it is cumbersome to classically simulate a stabilizer state directly, as one requires exponentially long state vectors encoding the physical space, to cover the vectors on a comparatively small manifold. It is therefore the goal of this section to strip such systems of their stabilizers, which removes all qubits down to the computational state, eliminating a considerable overhead for classical simulation.\\

\noindent\textbf{Notation}:$\;$ $X$, $Y$ and $Z$ are the three Pauli matrices and $\id$ is the identity. Whenever we deem it necessary, we will denote operators and states with subscripts, indicating the qubits they live on: $X_{3}$ for instance is a Pauli matrix acting on qubit number three. For multi-qubit states and operators, the subscript will include a set of qubits, for instance $S_{[a,b]}$, where $S$ acts on the qubits $[a,b] = \lbrace a, a+1, \dots, b \rbrace$. A similar shorthand excludes the first qubit, $\left(a, b\right] = \lbrace a+1, a+2, \dots, b\rbrace$. \\

The idea is to bring every Pauli string of the physical system ($n$ qubits, $r$ stabilizer generators) into the form
\begin{align}
    \label{eq:tapering}
    p^{\phantom{p}}_{[1,n]} \;  \mapsto \; \tau^p_{[1,r]} \otimes \Lambda^p_{(r,n]} \, ,
\end{align}
where $\tau^p$ is a Pauli string on the first $r$ qubits and $\Lambda^p$ is a Pauli string with a phase on the other $n-r$ qubits, such that the substrings $\tau^p$ of all logical operators $p$  commute with each other. We would then eliminate stabilizer conditions by applying  \eqref{eq:tapering} to all Pauli strings $p$ of a system, and discard the first $r$ qubits deleting the respective substrings $\tau^p$. For an operator with coefficients $\alpha(p)$ that means
\begin{align}
    \label{eq:reduce}
    \sum_p \alpha(p) \, p^{\phantom{p}}_{[1,n]} \; \mapsto \; \sum_p \alpha(p) \, \Lambda^{p}_{[1,n-r]} \, .
\end{align}
We can make the case that the remaining $\Lambda^p$ are the logical representations of the strings $p$, the operator actions of $p$  on the computational subspace of the quantum code. We will now present an explicit algorithm for \eqref{eq:tapering}, that is unique to a list of stabilizer generators. The algorithm is comprised of iterations of a protocol that takes a single stabilizer $S$ of an $m$-qubit system and transforms all its Pauli strings such that the Pauli operator on a fixed qubit  is rendered trivial in its action on the stabilizer state, then removes that qubit along with the stabilizer condition. 

We start by selecting a qubit on which $S$ has a nontrivial Pauli operator $u$ (which is $X$, $Y$ or $Z$, but not the identity $\id$). For the sake of clarity, we change the qubit order such that the selected qubit carries the label $1$. The stabilizer has the form
\begin{align}
    \label{eq:stabilizer}
    S^{\phantom{p}}_{[1,m]} = u^{\phantom{p}}_1 \otimes t^{\phantom{p}}_{(1,m]} \, ,\quad  u \in \lbrace X,Y,Z\rbrace \, .
\end{align}
The algorithm now matches $u$ with the Pauli operators $v$ and $w$ such that $(u,v,w)$ is some permutation of $(X,Y,Z)$. Let us denote the $+1$ eigenstate of $w$ by  $|w\rangle$. Any $m$-qubit state $| \psi \rangle$ stabilized by $S$ can be described by some $(m-1)$-qubit state $|\phi\rangle$ as
\begin{align}
    \label{eq:ansatz}
    \left| \psi \right\rangle_{[1,m]}^{\phantom{p}} = \frac{1}{\sqrt{2}}\left(\id^{\phantom{p}}_{[1,m]}+ u^{\phantom{p}}_1 \otimes t^{\phantom{p}}_{(1,m]}\right) \left| w \right\rangle_{1}^{\phantom{p}}\otimes\left| \phi \right\rangle_{(1,m]}^{\phantom{p}} \, .
\end{align}
Using this ansatz, we can completely represent $| \psi \rangle$ by  $| \phi \rangle$ if we can find the signed $(m-1)$-qubit Pauli string $p^{\prime}$ to every $m$-qubit Pauli string $p^{\vphantom{\prime}}$ such that $p|\psi\rangle$ can be represented by $p^\prime | \phi\rangle$. Let us assume that $p$ looks like
\begin{align}
    \label{eq:fixing0}
    p^{\phantom{p}}_{[1,m]} = \sigma^{\phantom{p}}_{1} \otimes q^{\phantom{p}}_{(1,m]} \,.
\end{align}
By multiplying $p$ with $S$ whenever $\sigma$ is not $w$ or the identity, the action of the resulting string is trivial on $|w\rangle$, and we find 
\begin{align}
     p_{[1,m]}\left| \psi \right\rangle_{[1,m]}^{\phantom{p}} = \frac{1}{\sqrt{2}}\left(\id^{\phantom{p}}_{[1,m]}+ u^{\phantom{p}}_1 \otimes t^{\phantom{p}}_{(1,m]}\right) \left| w \right\rangle_{1}^{\phantom{p}}\otimes p_{(1,m]}^\prime\left| \phi \right\rangle_{(1,m]}^{\phantom{p}}
\end{align}
where
\begin{align}
    \label{eq:fixing1}
    p^{\prime} = \left\lbrace \begin{tabular}{rl}
           $q \cdot t$&\quad if  $\sigma = u$  \\
           $\; -isq\cdot t$ &\quad if $\sigma = v$ \\
           $q$ &\quad  else,  
    \end{tabular} 
  \right .
\end{align}
where $s = \mathrm{sign}(iu\cdot v \cdot w)$ is the sign of the permutations that map $(X,Y,Z)$ to $(u,v,w)$.

So far we have assumed that  $p$ is a logical operator such that $p$ and $S$ commute. However, similar rules apply in the case where $S$ and $p$ anticommute. Reusing notation \eqref{eq:fixing0} for an anticommuting $p$, we find

\begin{align}
    \label{eq:fixing2}
    p^{\prime} = \left\lbrace \begin{tabular}{rl}
           $-q \cdot t$&\quad if  $\sigma = u$  \\
           $\; isq\cdot t$ &\quad if $\sigma = v$ \\
           $q$ &\quad  else
    \end{tabular} 
  \right .
\end{align}
in place of \eqref{eq:fixing1}. This is the same transformation with a flipped stabilizer $S \mapsto -S$. Indeed the state is suddenly stabilized by $-S$ rather than $S$
\begin{align}
    \label{eq:reduced}
   p \left| \psi \right\rangle_{[1,m]}^{\phantom{p}} = \frac{1}{\sqrt{2}}\left(\id^{\phantom{p}}_{[1,m]}- S^{\phantom{p}}_{[1,m]}\right) \left| w \right\rangle_{1}^{\phantom{p}}\otimes p^\prime\left| \phi\right\rangle_{(1,m]}^{\phantom{p}}  .
\end{align}
This state is no longer in the code space, as it is now stabilized by $-S$.  To distinguish the logical state $p|\phi\rangle$ stabilized by $-S$ from the logical state $p|\phi\rangle$ stabilized by $S$, we need to track the signs of the stabilizer generators, also called syndromes. States of different syndromes are always orthogonal to each other.

Storing states $|\phi\rangle$ and operators  $p^\prime$ (together with possible syndrome changes), effectively removes one qubit from the system and we lose one stabilizer condition:  $S$  has no meaning for $|\phi\rangle$, in fact $S^\prime = \id$. Let us now iterate this procedure starting with $m=n$ qubits and a list of $r$ stabilizer generators $(S^{(1)}, S^{(2)},\, ...\,  , S^{(r)})$. In the first round, we select the first generator and reduce the entire system by one qubit -- even the stabilizer generators  $ S^{(2)},\, ...\,  , S^{(r)}$. The first stabilizer is discarded and every new round of the procedure begins with selecting the next stabilizer along with one of the remaining qubits until no generators are left and $r$ qubits are removed. For every Pauli string that might have created a syndrome, a bit string of length $r$ has to be stored, where the $m$-th bit being set typically indicates a syndrome on the $m$-th stabilizer generator. 

Note however that this generator set is not necessarily the initial set of generators $(S^{(1)}, S^{(2)},\, ...\,  , S^{(r)})$, as the stabilizers have been multiplied with one another. After the first round, $S^{(2)\prime}$ could be $S^{(2)} \cdot S^{(1)}$ where the first qubit has been removed. The second stabilizer in the new stabilizer set is then $S^{(2)} \cdot S^{(1)}$. Fortunately, these relationships are easy to keep track of.

We are now storing a $(n-r)$-qubit state $|\phi\rangle$, plus a string of $r$ bits signifying its syndrome pattern of some known set of generators. Given that we have selected some Pauli operators $u^{(k)}$, $w^{(k)}$ in the $k$-th round of the protocol, the relationship between the physical states $|\psi\rangle$ and logical states $|\phi\rangle$ in the code space is
\begin{align}
    \label{eq:stabstate}
    |\psi\rangle^{\phantom{\dagger}}_{[1,n]}\;=\;\left[ \frac{1}{2^{r/2}}\prod_{m=1}^{r} \left( \id^{\phantom{p}}_{[1,n]} +  {S}_{[1,n]}^{(m)} \right)\right] \; | {w^{(1)}} \rangle^{\phantom{p}}_{1} \otimes | {w^{(2)}} \rangle^{\phantom{p}}_{2} \otimes \cdots  \otimes | {w^{(r)}} \rangle^{\phantom{p}}_{r} \otimes  \left| \phi \right\rangle^{\phantom{p}}_{(r,n]} \, ,
\end{align}
and the first $r$ qubits of every Pauli string $p$  are fixed to $\tau^p \in \bigotimes_{k=1}^{r} \lbrace \id, w^{(k)} \rbrace$, and they act trivially on \eqref{eq:stabstate}, as demanded in \eqref{eq:tapering}. For different encounters of $u\in \lbrace  X,Y,Z\rbrace$, we choose to define the triples $(u,v,w)$ as $(X,Y,Z)$, $(Y,X,Z)$ and $(Z,Y,X)$ in an implementation of the described procedure in \textbf{Algorithm$\,$1}. Note that this algorithm is not new. Stabilizer-like symmetries have been exploited to remove qubits in a similar manner in prior arts \cite{bravyi2017tapering,setia2020reducing}.
\begin{table}[]
    \centering
    \begin{tikzpicture}
    \node[] (table) {
\begin{tabular}{l}
\textbf{Algorithm 1} (list of Pauli strings, list of stabilizer generators)\\
\begin{tabular}{r|l}
    \codeline{1}&$\;{\color{magenta}r}\vphantom{\frac{\sum}{\sum}} \leftarrow$ number of stabilizer generators \\
   $\quad$ \codeline{2} &$\;$repeat {\color{magenta}$r$} times: \\
    &\begin{tabular}{r|l}
         $\qquad$\codeline{3} &$\;{\color{magenta}S} \leftarrow\vphantom{\frac{\sum}{\sum}}$ first stabilizer from generator list  \\
         \codeline{4}&$\;$remove ${\color{magenta}S}\vphantom{\frac{\sum}{\sum}}$ from generator list\\
        \codeline{6}&$\;{\color{magenta}\mathtt{index}}\vphantom{\frac{\sum}{\sum}}$, ${\color{magenta}\mathtt{type}}$ 
        $\leftarrow$ position and type of first nontrivial Pauli operator in ${\color{magenta}S}$\\
        \codeline{7}& $\;$for $\forall {\color{magenta}p}$ in $\lbrace \vphantom{\frac{\sum}{\sum}}$stabilizer generators, Pauli strings$\rbrace\vphantom{\frac{\sum}{\sum}}$: \\ 
        &\begin{tabular}{r|l}
            $\qquad$\codeline{8} &$\;{\color{magenta}\mathtt{type}^\prime}\leftarrow$ type of Pauli operator on position ${\color{magenta}\mathtt{index}}$ in ${\color{magenta}p}$ \\ \\
            \begin{tabular}{r}\codeline{9}\! \\ \codeline{10}\!\! \\  \codeline{11}\!\!
 \\ \end{tabular}&\begin{tabular}{ll}
                $\;$if & $({\color{magenta}\mathtt{type}} = {\color{magenta}\mathtt{type}^\prime})$ or \\
                 & $({\color{magenta}\mathtt{type}} \in \lbrace X,Z \rbrace$ and $ {\color{magenta}\mathtt{type}^\prime} = Y)$ or \\ 
                 & $({\color{magenta}\mathtt{type}}  = Y$ and $ {\color{magenta}\mathtt{type}^\prime} = X)$:
            \end{tabular} \\
            &\begin{tabular}{r|l}
                \\
                $\qquad$ \codeline{12}& $\;{\color{magenta}p} \leftarrow {\color{magenta}p}\cdot {\color{magenta}S}$ \\
                \\
            \end{tabular}\\ 
            \codeline{13}&$\;{\color{magenta}p} \leftarrow$ remove qubit  ${\color{magenta}\mathtt{index}}$ and its Pauli operator in ${\color{magenta}p}\vphantom{\frac{\sum}{\sum}}$ \\
            \codeline{14}&$\;$update ${\color{magenta} p}\vphantom{\frac{\sum}{\sum}}$ in its source list
        \end{tabular} \\
        
    \end{tabular} \\ \\
    \codeline{15}&$\;$return list of Pauli strings $\vphantom{\frac{\sum}{\sum}}$
\end{tabular}
\end{tabular}};
    \fill [black, rounded corners=1em, opacity = .04] (table.north west) rectangle (table.south east);
\end{tikzpicture}

    \caption*{Algorithm 1: Given a list of stabilizer generators, this algorithm obtains the logical representation of a list of Pauli strings.}
    \label{tab:algo1}
\end{table}

\textbf{Algorithm 1} can also be modified to cure qubit Hamiltonians from a proliferation in their one-norm. Mappings like \eqref{eq:map1} and \eqref{eq:mapping} might increase the number of terms in the Hamiltonian, which is detrimental for quantum simulation. Fortunately, this increase can be reversed with a classical routine outlined in Appendix \ref{sec:n_terms}.
\section{Simulator} \label{simulator}
In this section, we will introduce a classical simulator verifying Claim \ref{clm:simulator}. We do this in two steps: first, we are introducing an ansatz for the density matrix of the system in Section \ref{subsec:density}, where we show that the ansatz is indeed consistent with the circuit under noise. Second, we provide the actual algorithm to reconstruct the aforementioned density matrix within Section \ref{subsec:algorithm}.
\subsection{Density matrix and noise}
\label{subsec:density}
We can simulate the time evolution of a stabilized system like in \eqref{eq:stabstate} under Pauli noise with the ansatz:

\begin{align}
    \label{eq:density_mtx}
    \varrho_{[1,n]}^{\phantom{\dagger}} = \sum_{\lambda\in \lbrace 0,1\rbrace^{r}} \frac{1}{2^r} \left[ \prod_{j = 1}^{r}\left( \id_{[1,n]}^{\phantom{\dagger}} + (-1)^{\lambda_j}S_{[1,n]}^{(j)}\right)\right] \left(\bigotimes_{k=1}^r  {|{w^{(k)}}\rangle\!\langle {w^{(k)}}|}^{\phantom{\dagger}}_{k} \right)\otimes {\rho^{\phantom{\dagger}}_{\lambda}}_{(r,n]} \left[ \prod_{l = 1}^{r}\left( \id_{[1,n]}^{\phantom{\dagger}} + (-1)^{\lambda_l}S_{[1,n]}^{(l)}\right)\right] ,  
\end{align}
where $\rho^{\phantom{\dagger}}_\lambda$ is a subnormalized density matrix  for the syndrome $\lambda = (\lambda_1, \lambda_2, ... \, , \lambda_r)$, with $\lambda_j \in \lbrace 0,1\rbrace$ for all $j$. Since two stabilizer states with different syndromes $\lambda$ are orthogonal, we can emulate \eqref{eq:density_mtx} by some density matrix
\begin{align}
    \label{eq:blocks}
     \rho_{[1,n]}^{\phantom{\dagger}} =\sum_{\lambda\in \lbrace 0,1\rbrace^{r}} {\rho^{\phantom{\dagger}}_{\lambda}}^{\phantom{\dagger}}_{[1,n-r]} \otimes |\lambda \rangle\!\langle \lambda |^{\phantom{\dagger}}_{(n-r,n]}\, ,
\end{align}
\begin{figure}
    \centering
    \begin{tikzpicture}
\draw[color=white, ultra thick, fill=black!10] (0,0) -- ++(1,0) -- ++(0,-1) -- ++(-1,0) -- cycle;
\draw[color=white, ultra thick, fill=black!10] (1,0) -- ++(1,0) -- ++(0,-1) -- ++(-1,0) -- cycle;
\draw[color=white, ultra thick, fill=black!10] (2,0) -- ++(1,0) -- ++(0,-1) -- ++(-1,0) -- cycle;
\draw[color=white, ultra thick, fill=orange] (3,0) -- ++(1,0) -- ++(0,-1) -- ++(-1,0) -- cycle;
\draw[color=white, ultra thick, fill=black!10] (0,1) -- ++(1,0) -- ++(0,-1) -- ++(-1,0) -- cycle;
\draw[color=white, ultra thick, fill=black!10] (1,1) -- ++(1,0) -- ++(0,-1) -- ++(-1,0) -- cycle;
\draw[color=white, ultra thick, fill=orange] (2,1) -- ++(1,0) -- ++(0,-1) -- ++(-1,0) -- cycle;
\draw[color=white, ultra thick, fill=black!10] (3,1) -- ++(1,0) -- ++(0,-1) -- ++(-1,0) -- cycle;
\draw[color=white, ultra thick, fill=black!10] (0,2) -- ++(1,0) -- ++(0,-1) -- ++(-1,0) -- cycle;
\draw[color=white, ultra thick, fill=orange] (1,2) -- ++(1,0) -- ++(0,-1) -- ++(-1,0) -- cycle;
\draw[color=white, ultra thick, fill=black!10] (2,2) -- ++(1,0) -- ++(0,-1) -- ++(-1,0) -- cycle;
\draw[color=white, ultra thick, fill=black!10] (3,2) -- ++(1,0) -- ++(0,-1) -- ++(-1,0) -- cycle;
\draw[color=white, ultra thick, fill=cyan] (0,3) -- ++(1,0) -- ++(0,-1) -- ++(-1,0) -- cycle;
\draw[color=white, ultra thick, fill=black!10] (1,3) -- ++(1,0) -- ++(0,-1) -- ++(-1,0) -- cycle;
\draw[color=white, ultra thick, fill=black!10] (2,3) -- ++(1,0) -- ++(0,-1) -- ++(-1,0) -- cycle;
\draw[color=white, ultra thick, fill=black!10] (3,3) -- ++(1,0) -- ++(0,-1) -- ++(-1,0) -- cycle;
\node[white] at (.5,2.55) {$\rho^{\phantom{\dagger}}_{00}$};
\node[white] at (1.5,1.55) {$\rho^{\phantom{\dagger}}_{01}$};
\node[white] at (2.5,.55) {$\rho^{\phantom{\dagger}}_{10}$};
\node[white] at (3.5,-.45) {$\rho^{\phantom{\dagger}}_{11}$};
\node[black!30] at (0.5,-.5) {0};
\node[black!30] at (1.5,-.5) {0};
\node[black!30] at (2.5,-.5) {0};
\node[black!30] at (0.5,.5) {0};
\node[black!30] at (1.5,.5) {0};
\node[black!30] at (3.5,.5) {0};
\node[black!30] at (0.5,1.5) {0};
\node[black!30] at (2.5,1.5) {0};
\node[black!30] at (3.5,1.5) {0};
\node[black!30] at (1.5,2.5) {0};
\node[black!30] at (2.5,2.5) {0};
\node[black!30] at (3.5,2.5) {0};
\draw[rounded corners, very thick, white] (0,-1) -- ++(4,0) -- ++(0,4) -- ++(-4,0) -- cycle; 
\end{tikzpicture}
    \caption{The density matrix of \eqref{eq:blocks}, for $r=2$ stabilizer generators. The density matrix is only populated in diagonal blocks of size $2^{n-r}\times2^{n-r}$, where each block contains the computational space representation of states with the same certain syndrome pattern. Amongst them,  we find the code space with a trivial syndrome pattern $\lambda = 00$ (cyan), and syndrome spaces with nontrivial syndrome patterns $01$, $10$ and $11$  (orange). The rest of the matrix is zero. The measurement of all stabilizer generators would project us into a block along the diagonal, and as nontrivial syndromes are flags for errors we would discard any quantum circuit evaluation that does not land us in the code space. }
    \label{fig:blocks}
\end{figure}
where the density matrix is block diagonal, and each block $\rho_\lambda$ corresponds to the logical state with respect to a syndrome pattern $\lambda$, see Figure \ref{fig:blocks}. 
Without noise, we would find  $\rho^{\phantom{\dagger}}_{0^{r}} = |\phi\rangle \!\langle\phi| $ for some pure state $|\phi\rangle$ and $\rho^{\phantom{\dagger}}_{\lambda} = 0$ for all other syndrome patterns $\lambda \neq 0^r$. We will now show that the system can still be described by \eqref{eq:blocks} when exposed to Pauli noise during and in between quantum subroutines.
The Pauli noise can be described by a set $\lbrace (\eta_j,P_j)\rbrace_j$ of probabilities $\eta_j$ such that $\sum_j \eta_j = 1$ and Pauli strings $P_j \in \lbrace X,Y,Z,\id \rbrace^{\otimes n} $. The noise channel takes the form:
\begin{align}
    \label{eq:idle}
   \varrho^{\phantom{\dagger}}_{[1,n]} \; \mapsto \; \sum_{j} \eta_j \, {P^{\phantom{\dagger}}_{j}}_{[1,n]} \, \varrho^{\phantom{\dagger}}_{[1,n]} \, {P^{\phantom{\dagger}}_j}_{[1,n]} \, .
    \end{align}
Using \eqref{eq:tapering}, we find that the channel maps 

\begin{align}
    \label{eq:idle2}
    \rho \; \mapsto \; \sum_{j} \eta_j\, \Lambda^{P_j}_{[1,n-r]} \,  {\rho_{\lambda}^{\phantom{\dagger}}}_{[1,n-r]} \, \left(\Lambda^{P_j}_{[1,n-r]}\right)^\dagger \otimes \, |\lambda\oplus\nu_j\rangle\!\langle \lambda\oplus\nu_j |^{\phantom{\dagger}}_{(n-r,n]} \, ,
\end{align}
 where $\oplus$ is the vector addition modulo 2. In \eqref{eq:idle2}, $P_j$ has caused a syndrome pattern $\nu_j = (\nu_{j,1}, \nu_{j,2}, ... \, , \nu_{j,r})$, as $P_j$ anticommutes with those Pauli strings $S^{(m)}$ for which $\nu_{j,m}=1$.
We can now show that Pauli noise conserves the ansatz by reformulating \eqref{eq:idle2} as an update of the density matrix blocks \begin{align}
     \rho^{\phantom{\dagger}}_{\lambda} \; \mapsto \; \sum_{j}\eta_j\, \Lambda^{P_j} \rho^{\phantom{\dagger}}_{\lambda\oplus\nu_j}\left(\Lambda^{P_j}\right)^\dagger\, .
 \end{align}
We now have to consider an individual tapering process for each syndrome pattern $\lambda$: depending on whether a Pauli string $p$ commutes with a stabilizer or not, its logical representation is either \eqref{eq:fixing1} or \eqref{eq:fixing2}, which in the end can lend a minus sign to the operator $\Lambda^p$.  We therefore denote the logical representation of the Pauli string $p$ on the state with syndrome pattern $\lambda$ by $\Lambda^{p,\,\lambda}$. In \eqref{eq:idle2}, this effect has been ignored as the signs would cancel. It does make a difference however for Pauli string rotations of logical operators $p$: $\exp(i\theta p)$. A noise-free Pauli string  rotation around the angle $\theta$ corresponds to  updating the density matrix blocks as 
\begin{align}
    \label{eq:rotation}
     \rho^{\phantom{\dagger}}_{\lambda} \; \mapsto \; \exp\!\left(\vphantom{\sum}i\theta \,\Lambda^{p, \, \lambda}\right) \rho^{\phantom{\dagger}}_{\lambda}\,\exp\!\left(-i\vphantom{\sum}\theta \Lambda^{p, \, \lambda}\right)\, .
 \end{align}
 We now have all the ingredients to show that the ansatz remains valid even when the circuits are noisy. The type of circuits with which we would simulate a Pauli string rotation is depicted in Figure \ref{fig:circuit}(a). Apart from the $Z$-rotation in the center, the circuit only consists of Clifford gates and so any statistically-placed Pauli operator can be enclosed within a Clifford circuit such that it falls out of the rotation gadget, see Figure \ref{fig:circuit}$(b)$,$(c)$. As Clifford circuits normalize the Pauli group, the circuit might propagate Pauli noise operators, but they remain Pauli, as is concluded in Figure \ref{fig:circuit}$(d)$. With Pauli  noise being pushed out of the rotation circuits, we can show that the system conserves the ansatz state \eqref{eq:ansatz} by concatenation of \eqref{eq:idle2} and \eqref{eq:rotation}. The system therefore only propagates inside the block diagonals of \eqref{eq:blocks}, and we can envision a vector simulator that would allow us to reconstruct the blocks  $\rho^{\vphantom{p}}_{\lambda}$ by simulating the logical rather than physical space of the system.
 \subsection{Simulation algorithm}
 \label{subsec:algorithm}
 We will now sketch an algorithm that can populate the density matrix blocks $\rho_\lambda$ with state vector samples. 
 As rotations and noise need to pass into the logical subspace by \textbf{Algorithm$\,$1}, the envisioned simulator needs to be part numerical and part symbolical. 
 \begin{figure}
    \centering
    \includegraphics[scale=.4]{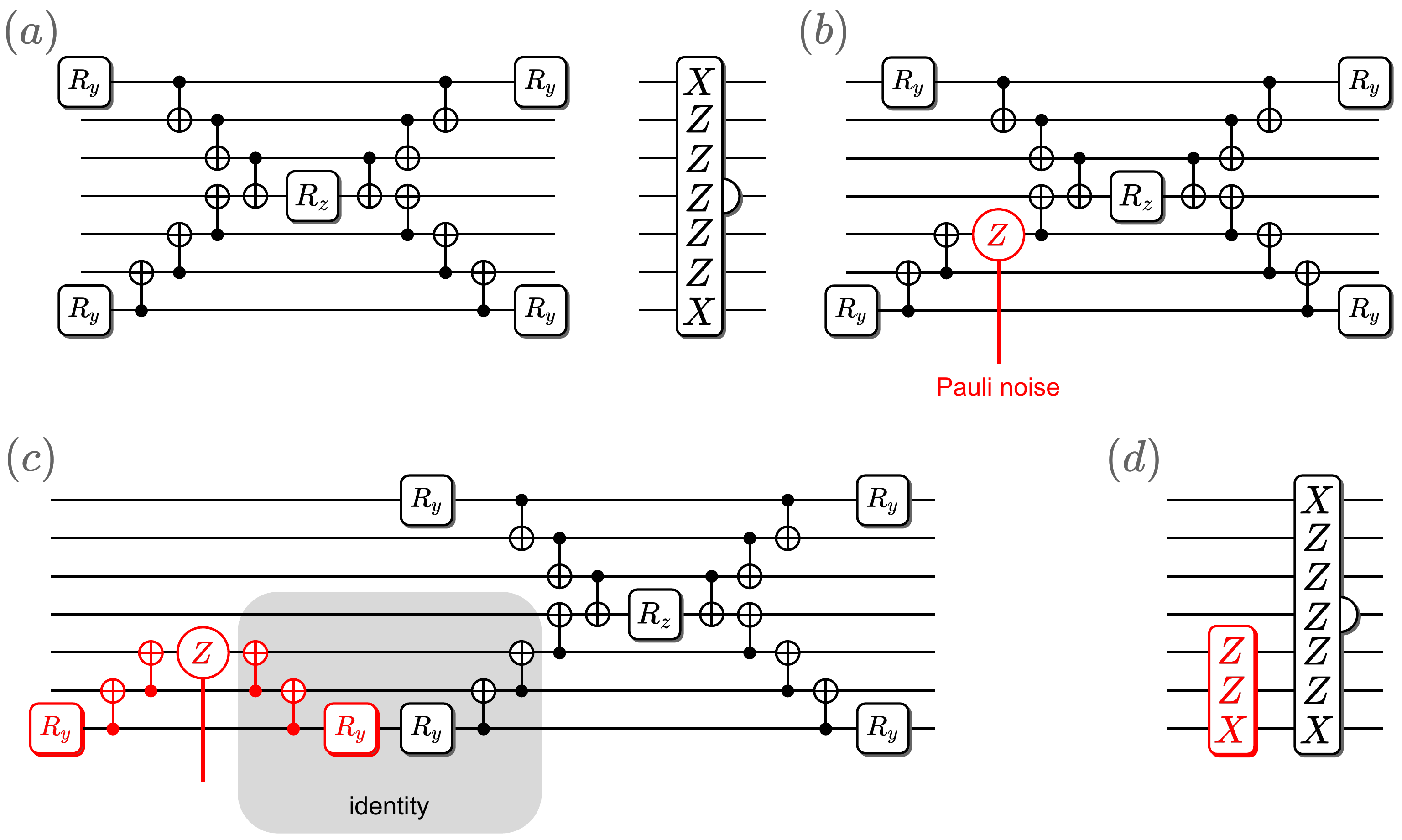}
    \caption{Evolution of a Pauli string rotation subcircuit under noise. $(a)$ Noiseless rotation of the Pauli string $X \otimes Z^{\otimes 5} \otimes X$ as a composite gate (right) and decomposed into elementary gates (left). $R_y$ and $R_z$ are $Y$ and $Z$ Pauli rotations, respectively. $(b)$ Randomly inserting a Pauli noise operator (in this case a $Z$ operator, red) into the rotation circuit. $(c)$ Inserting a noiseless Clifford circuit equal to the identity (gray indicated area) encapsulating the noise operator into a Clifford circuit (red). $(d)$ The dressed noise operator (red) turns into a Pauli string separate from the rotation. }
    \label{fig:circuit}
\end{figure}
With a vector simulator, we would approximate the density matrix in shots, where each shot is a complete execution of the entire circuit with statistically placed noise operators: we randomly place Pauli operators into all the subroutines for Pauli rotations and the idle periods between them. Let us first zoom into the Pauli rotation subroutines.

When considering a circuit such as in Figure \ref{fig:circuit}(a), we would begin by expressing the rotation $R_z(\theta)$ in its symbolical form $\cos\theta + i\sin\theta Z$, and then subject this expression to the two $\mathrm{CNOT}$ gates on the left- and right-hand side of $R_z$ in the circuit. One could for instance just multiply $R_z$ with the  symbolic expressions of the $\mathrm{CNOT}$ gates.  Ideally, the resulting expression would be $\cos\theta + i\sin\theta Z\otimes Z$, but we might have inserted Pauli operators into the calculation at random, in order to account for noise. These Pauli operators $P_j$ are one- or two-qubit Kraus operators, randomly drawn from the statistical or gate noise model according to their respective probabilities $\eta_j$. Gate noise operators have a chance to be placed after a two-qubit gate while statistical noise operators must be considered in any time step on any quantum wire. A noise operator $X\otimes X$ right after the first $\mathrm{CNOT}$ would for instance lead to a symbolic expression of
 \begin{align}
   \mathrm{CNOT}\;(\id \otimes R_z(\theta))(X\otimes X)\,\mathrm{CNOT} =   \cos\theta\; (X \otimes \id)  - \sin\theta\; (Y\otimes Z)\, .
 \end{align}
 
 We develop the expression further by sandwiching it with the next two $\mathrm{CNOT}$ gates in the circuit, always considering to place noise operators after gates and on wires. Continuing this process with one pair of $\mathrm{CNOT}$ gates after the other, we are slowly encompassing the entire circuit from the inside out. By choosing every pair of gates to be mirror images with respect to a vertical axis through the $R_z$ rotation in the circuit, we guarantee that the symbolical expression has only two terms. The process terminates after the symbolical expression is sandwiched by the last $R_y$ rotation in the circuit. We have obtained a unique snapshot of the noisy Pauli string rotation in symbolical form.\\
 
\begin{figure}
    \centering
    \includegraphics[scale=.5]{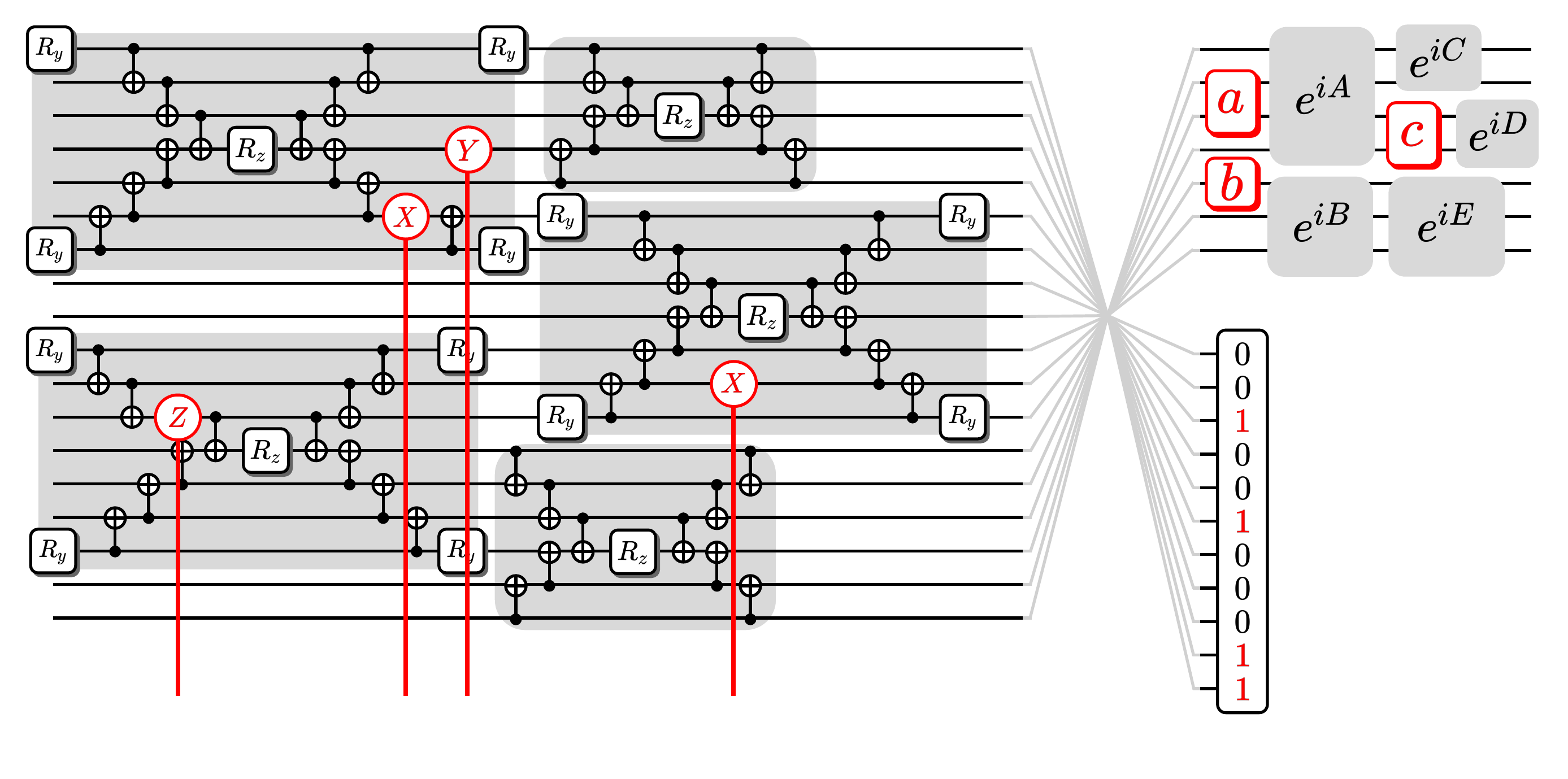}
    \caption{A single shot of the simulator. \textbf{(Left)}: Rotation subroutines (gray) represent events of a single circuit evaluation, in which noise operators (red) have been inserted at random. \textbf{(Right)}: The same circuit is transformed by \textbf{Algorithm\,2} into two different evaluations. (1: above) A logical subspace representation on fewer qubits, where the rotation circuits turn into the logical operators $e^{iA}, e^{iB}, ..., \,  e^{iE}$, and Pauli errors separate from the rotations as Pauli strings $a$, $b$ and $c$. (2: below) A unique syndrome pattern acquired by the errors. The number of logical qubits plus the length of the syndrome string is equal to the number of physical qubits on the left.}
    \label{fig:shot}
\end{figure}

A set of such snapshots for all rotations in the circuit, plus snapshots of noise during idle periods in between the circuits make up the list of events $(U_{(1)}, U_{(2)}, \, ...\, U_{(M)})$ for a single shot. In every shot, we use \textbf{ Algorithm$\,$1} to map the algorithm to operators acting onto a $(n-r)$-qubit state vector and a syndrome pattern, see Figure \ref{fig:shot}.
We start with a dense vector of the initial state $| \phi_{(0)}\rangle$ and a length-$r$ bit string signifying the initial syndrome pattern  $\lambda_{(0)}$. These two quantities would come from the state preparation routine outlined in Appendix \ref{sec:stateprep}. The first event $U_{(1)}$ in the circuit could potentially flip some stabilizers  and so we update the syndrome pattern accordingly to $\lambda_{(0)} \mapsto\lambda_{(1)}$. We can now process the first event's impact on the computational state by updating  $|\phi_{(0)}\rangle \mapsto|\phi_{(1)}\rangle = \Lambda^{U_{(1)}, \, \lambda_{(1)}}|\phi\rangle$, where $\Lambda^{U_{(1)}, \, \lambda_{(1)}}$ has been converted into a matrix. The first event is now processed and $U_{(2)}$ acts on $\lambda_{(1)}$ and $|\phi_{(1)}\rangle$ in the second round of this procedure. The $m$-th round begins by checking whether one of the Pauli strings in $U_{(m)}$ would anticommute with stabilizer generators and flip the corresponding bits in the syndrome pattern $\lambda_{(m-1)}\mapsto\lambda_{(m)}$, before updating $|\phi_{(m-1)}\rangle \mapsto |\phi_{(m)}\rangle$ with 
\begin{align}
    \left|\phi_{(m)}\right\rangle =  \Lambda^{U_{(m)}, \, \lambda_{(m)}}\left|\phi_{(m-1)}\right\rangle \, .
\end{align}
The result of the last round, $|\phi_{(M)}\rangle$ is  transformed into a density matrix $|\phi_{(M)}\rangle\!\langle\phi_{(M)}|$ and recombined with the block $\rho_{\lambda_{(M)}}$. The entire procedure is summarized in \textbf{Algorithm$\,$2}. 

Note that we are not necessarily interested in keeping all the blocks around. In postselection for instance, we would measure all stabilizer generators and discard a state if we detected a syndrome $\lambda \neq 0^r$, and so we can discard any shots that are not in the code space. Whenever truncating blocks, one must normalize the remaining density matrix block(s) with the number of binned shots, rather than the total number of shots. The ratio of total and binned number of shots can be an interesting quantity, as it describes the simulation overhead in postselection. In the next section, we are demonstrating \textbf{Algorithm 2} numerically using the \simulator simulator developed in conjunction with this work. The \simulator simulator is based on OpenFermion, using the packages' symbolical representation of Pauli strings. A source for the simulator as well as instructions for its use can be found at 
\begin{align*}\mathtt{https://github.com/msteudtner/pylove\_simulator} \, .
\end{align*}

\begin{table}[]
    \centering
 \begin{tikzpicture} 

 \node[] (table) {\begin{tabular}{l}
\textbf{Algorithm$\,$2} (list of stabilizer generators, list of operators for state preparation, list of rotations, noise model, number of shots)$\vphantom{\frac{\sum}{\sum}}$\\ 
\begin{tabular}{r|l}
$\qquad$ \codeline{1}& $\;{\color{magenta} \mathtt{schedule}} \leftarrow\vphantom{\frac{\sum}{\sum}}$ make list of circuit events, that is idle periods and rotations \\
\codeline{2}&$\;${\color{magenta} $\rho$} $\leftarrow\vphantom{\frac{\sum}{\sum}}$ container of empty density matrix blocks\\
\codeline{3}&$\;${\color{magenta} $n$} $\leftarrow\vphantom{\frac{\sum}{\sum}}$ number of physical qubits\\
\codeline{4}&$\;${\color{magenta} $r$}  $\leftarrow\vphantom{\frac{\sum}{\sum}}$ number of stabilizer generators\\
\codeline{5}&$\;$repeat for every shot: \\
&   \begin{tabular}{r|l}

    $\qquad$ \codeline{6}&$\;{\color{magenta} \mathtt{syndr}} \leftarrow$  empty bit string of length $\color{magenta}r\vphantom{\frac{\sum}{\sum}}$, denoting syndromes \\
            \codeline{7}& $\;{\color{magenta} \mathtt{gen}\vphantom{\frac{\sum}{\sum}}} \leftarrow $ empty container for ${\color{magenta} r}$ stabilizer generators\\
            \codeline{8}& $\;{\color{magenta} \mathtt{state}} \leftarrow $  container for dense state vector on ${\color{magenta}n}-{\color{magenta}r}\vphantom{\frac{\sum}{\sum}}$ qubits \\
                $\qquad$ \codeline{9}&$\;{\color{magenta} \mathtt{state}}, \vphantom{\frac{\sum}{\sum}}{\color{magenta} \mathtt{gen}}, {\color{magenta} \mathtt{syndr}}\leftarrow$ dense state vector, stabilizer generators, syndromes from initial state preparation\\
            \codeline{10}&$\;$for ${\color{magenta} \mathtt{event}}$ in ${\color{magenta} \mathtt{schedule}\vphantom{\frac{\sum}{\sum}}}$:\\
&   \begin{tabular}{r|l}
        $\qquad$ \codeline{11}&$\;{\color{magenta} U} \leftarrow \vphantom{\frac{\sum}{\sum}} $ get symbolical representation of ${\color{magenta} \mathtt{event}}$ after randomly placing noise operators into it \\
        \codeline{12}&$\;$for ${\color{magenta} i} \in [1,{\color{magenta} r}]:\vphantom{\frac{\sum}{\sum}}$ \\
        &   \begin{tabular}{l|l} \\
            $\qquad$ \codeline{13}&$\;$if ${\color{magenta} \mathtt{gen}_i}\vphantom{\frac{\sum}{\sum}}$ anticommutes with a Pauli string in ${\color{magenta} U}$:\\
            &   \begin{tabular}{r|l}
                $\qquad$\codeline{14} &$\;{\color{magenta} \mathtt{gen}^{\phantom{\dagger}}_i} \leftarrow (-1) \,  {\color{magenta} \mathtt{gen}^{\phantom{\dagger}}_i}\vphantom{\frac{\sum}{\sum}}$  \\
                \codeline{15}&$\;{\color{magenta} \mathtt{syndr}^{\phantom{\dagger}}_i} \leftarrow   {\color{magenta} \mathtt{syndr}^{\phantom{\dagger}}_i} + 1  \;\; \mathrm{mod}\; 2\vphantom{\frac{\sum}{\sum}}$  \\
                \end{tabular}\\
                \\
            \end{tabular}\\
            \codeline{16}&$\;{\color{magenta} \mathtt{ops}} \leftarrow $ \textbf{Algorithm$\,$1}$\,$(list of Pauli strings in ${\color{magenta} U}$, copy of ${\color{magenta} \mathtt{gen}}\vphantom{\frac{\sum}{\sum}}$)\\ 
            \codeline{17}&$\;{\color{magenta} \mathtt{ops}} \leftarrow $  recombine ${\color{magenta} \mathtt{ops}\vphantom{\frac{\sum}{\sum}}}$ into a single operator and convert it into a matrix \\
            \codeline{18}&$\;{\color{magenta} \mathtt{state}} \leftarrow {\color{magenta} \mathtt{ops}} \cdot {\color{magenta} \mathtt{state}}\vphantom{\frac{\sum}{\sum}}$
            
        \end{tabular}\\
        \codeline{19}&$\;{\color{magenta} \mathtt{state}} \leftarrow$ convert ${\color{magenta} \mathtt{state}}\vphantom{\frac{\sum}{\sum}}$ into a density matrix \\
       \codeline{20} &$\;{\color{magenta}\rho}^{\phantom{\dagger}}_{\left[{\color{magenta} \mathtt{syndr}}\right]} \leftarrow {\color{magenta} \rho}^{\phantom{\dagger}}_{\left[{\color{magenta} \mathtt{syndr}}\right]} + {\color{magenta} \mathtt{state}}\vphantom{\frac{\sum}{\sum}}$
    \end{tabular}\\ \\
    \codeline{21}&$\;{\color{magenta} \rho} \leftarrow$ normalize ${\color{magenta} \rho} \vphantom{\frac{\sum}{\sum}}$  \\ 
    \codeline{22}&$\;$return  ${\color{magenta} \rho} \vphantom{\frac{\sum}{\sum}}$
\end{tabular}
\end{tabular}};
    \fill [black, rounded corners=1em, opacity = .04] (table.north west) rectangle (table.south east);
\end{tikzpicture}
    \caption*{Algorithm 2: Reconstructs density matrix blocks of a system stabilized by a given list of generators. The state gets initialized with respect to a given set of Pauli strings and passes through a circuit described by a  list of Pauli string rotations. The initialization and the quantum circuit are both subject to a  Pauli noise model. The state is reconstructed using a certain number of shits, where Pauli noise operators are statistically inserted in between the scheduled circuit events in every shot. The stabilizer generators,  initialization, quantum circuit and noise model can be defined by the user.}
    \label{tab:algo2}
\end{table}

\section{Numerical experiments} \label{numerical_experiments}

\begin{table}
    \begin{tabular}{c|c|c|c|c|c|c|c|c}
        Encoding family & $[[n,k,d]]$ & $\mathrm{avg.\, wt}$ $\widehat{\mathcal{A}}$ & $\mathrm{avg.\, wt}$ $\widehat{\mathcal{B}}$  & $\mathrm{avg.\, wt}\, S $ & 9 mode rand. & 12 mode rand. & VQE & dynamics  \\
        \hline
        \rowcolor{black!8}JW & $[[9,9 ,1]]$ & 3 & 1 & N/A & {\color{black!40}\ding{51}} & & & \\
        Compact & $[[11,9 ,1]]$ & 2.67 & 1 & 6 & {\color{black!40}\ding{51}} & & &\\
        \rowcolor{black!8}GSE & $[[14,8 ,1]]$ & 2.71 & 1.56 & 4.67 & {\color{black!40}\ding{51}} & & &\\
        \hline
        JW1 & $[[12,12 ,1]]$ & 3.11 & 1 & N/A &  &{\color{black!40}\ding{51}}   & {\color{black!40}\ding{51}} &{\color{black!40}\ding{51}}\\
        \rowcolor{black!8}JW2 & $[[12,12 ,1]]$ & 3.41 & 1 & N/A &  &{\color{black!40}\ding{51}} & & \\
        Compact & $[[14,12,1]]$ & 2.57 & 1 & 5 &  & & {\color{black!40}\ding{51}} &\\
        \rowcolor{black!8}Compact & $[[15,12,1]]$ & 2.71 & 1 & 6.33 &  &{\color{black!40}\ding{51}}  & & {\color{black!40}\ding{51}} \\
        GSE & $[[16,10,1]]$ & 2.43 & 1.33 & 4 &  & & {\color{black!40}\ding{51}} & \\
        \rowcolor{black!8}GSE & $[[20,12,1]]$ & 2.76 & 1.67 & 5 &  & {\color{black!40}\ding{51}} &  & {\color{black!40}\ding{51}} \\
        GAQM & $[[25,12,2]]$ & 4.12 & 2.67 & 5.46 &  &{\color{black!40}\ding{51}}   & {\color{black!40}\ding{51}} &{\color{black!40}\ding{51}} \\
        \rowcolor{black!8}GSE & $[[34,11,2]]$ & 2.59 & 2.83 & 5.09&  &{\color{black!40}\ding{51}}   & &\\
        GSE$^*$ & $[[42,10,3]]$ & 4 & 3 & $7.8$ &  & & {\color{black!40}\ding{51}} &\\
    \end{tabular}
    \caption{Parameters for encodings used in the numerical experiments. The abbreviations used in the first column are as follows; JW: Jordan-Wigner transform, Compact: Compact encoding, GSE: Generalized Superfast encoding, GAQM: Generalized Auxiliary Qubit Encoding.  The parameters $[[n,k,d]]$ in the second column describe the error-correction capabilities of each code, where  $n$ is the number of physical qubits, $k$ is the number of logical qubits ($\log_2$ of the codespace dimension), and $d$ is the code distance (the minimum weight of all logical operators). The third column lists the  weight of the edge operators $\widehat{\mathcal{A}}$  averaged over all nearest-neighbor pairs. We also compare the average weights of $\mathcal{B}$ operators and stabilizer generators -- for all these operators we have chosen a form that minimizes their weight. The final four columns indicate which encodings are used in each simulation. The $[[42,10,3]]$ GSE code (decorated with a $^*$) is constructed around a pair of $7$-mode complete graphs. The parity in each of the two clusters is then included in the stabilizer group and so are $\mathcal{B}$ operator of the 7th modes in each graph in order to freeze out these modes and bring the number of encoded modes to 12. }
    \label{tab:simulation_encodings}
\end{table}

Next, we employ the \simulator simulator to investigate the  error-mitigation capabilities of several quantum codes numerically. That is, we run many shots of the same quantum circuits with statistically inserted noise, tracking  syndrome vectors and logical states. Density matrices are reconstructed and then evaluated. Within this investigation we are only interested in the codespace, and so we discard all shots with a syndrome pattern indicating errors.  When the experiment is very noisy or the code has a poor performance, the factor by which we must increase the number of shots in order to attain the same number of kept shots as the unmitigated simulation (a number we refer to as the sampling-increase rate) can be very high and so it can become challenging to get enough samples. This is particularly the case for mappings with larger stabilizer groups and deep circuits. We will make a note of instances where more shots would have helped  to converge the data points to their expected value, although we always take sufficiently many shots for the results to be reliable.

Note that these experiments do not necessarily advocate for measurement postselection in the systems. In fact, we want to keep these results agnostic of the error mitigation algorithm, and so we discount the errors that would occur within the measurement circuits in postselection. Our results are to be regarded as the codes general capacity for error mitigation in the respective experiment. In the best possible performance of an error mitigation algorithm,  all syndrome spaces but the codespace have been filtered out. An alternative to postselection is the quantum subspace expansion algorithm proposed in \cite{mcclean2020decoding}. 

The results for a particular code will depend on a number of factors including its circuit depth, the number of physical qubits,  code distances and numbers of stabilizer generators. Many times it is unclear whether some of these factors outweigh others, for instance whether trying to achieve a high code distance is a better strategy than trying to keep logical operators light.  Attempting a fair comparison of various codes, we investigate the performance of a number of existing mappings for a couple instances of different sizes (implying varying numbers of physical qubits).

To that end, our study includes three basic quantum simulation tasks: (1) random fermionic circuits where we investigate the fidelity of the error-mitigated state with the ideal state in the codespace as well as the sampling overhead, (2) optimization of a variational wavefunction (VQE) of a  6-site (12 mode) Fermi-Hubbard model on a $2\times 3$ grid, (3) simulation of a model of non-equilibrium discrete-Floquet dynamics featuring uni-directional edge modes. All these experiments are simulated with a pair of independent $1\%$ depolarizing noise channels acting after each $\mathrm{CNOT}$ gate on the two qubits involved.

Before discussing each of the three experiments in greater detail, we will briefly summarize the results. In all three of our experiments on systems containing $9$ and $12$ fermionic modes, we see that the use of a locality-preserving encoding in combination with stabilizer verification can increase the fidelity of an output quantum state compared to both the unmitigated output state and to the output state when encoded with the Jordan-Wigner transform, which uses no extra qubits and has a trivial stabilizer group.

In our random circuit experiments, the Compact encoding performs especially well. The fidelities of the error mitigated states are among the highest of all encodings. Utilizing the fewest additional qubits, it comes with the smallest nontrivial stabilizer group of any encoding at each system size. This immediately implies that the sampling overhead should also remain the lowest, which is borne out in our numerical data. 
Our VQE experiments have demonstrated an improved ability to optimize variational parameters when estimating the gradients with stabilizer postselection error mitigation.
Our numerical simulation of dynamics shows that combining a locality-preserving encoding with stabilizer postselection can meaningfully increase the fidelity of noisy quantum simulation of model dynamics.

\subsection{Random fermionic circuits}

Before discussing the error mitigation performance of quantum codes on problems involving structured circuits, we want to consider circuits with less structure. That is, we will consider circuits of logical rotations by randomly chosen angles with generators drawn uniformly at random from the set of edge, vertex and vertex-pair operators $\mathcal{A}_{ij}$, $\mathcal{B}_i$, $\mathcal{B}_i \mathcal{B}_j$ where all tuples $(i,j)$ denote geometrically-adjacent fermionic modes.  The locality of their qubit operators, however, depends on the code and may vary. The code parameters and average operator weights for the random circuit numerical experiments are presented in Table (\ref{tab:simulation_encodings}), where we have considered a $3\times 3$ square lattice of 9 fermionic modes and a $4\times 3$ rectangular lattice of 12 modes. 

Note that we deal with two types of randomness in our random circuit simulations that we average over. First, the randomness coming from the probabilistic noise model defining the density matrix and second, the randomness coming from the different random circuit realizations.  While the rotation angles are chosen at random, we perform the same logical-operator rotations for each encoding resulting in circuits of different depths for each mapping. For each randomly generated rotation sequence, we calculate the fidelity against the output pure state of same circuit run without noise and we count the number of shots in which the state was observed to lie in the codespace, providing an empirical estimate of the sampling increase factor. Each of these are then averaged over the different random circuit realizations. For the $9$ mode random circuit experiments, we average over $100$ different random circuit realizations. For the more computationally expensive $12$ mode experiments, we average over $10$ random circuit realizations.
Note that the circuits are not constructed in a layered fashion, the generator for each rotation is drawn at random from the set above.

\begin{figure}[htp]

\begin{subfigure}{\textwidth}
\includegraphics[width=\textwidth]{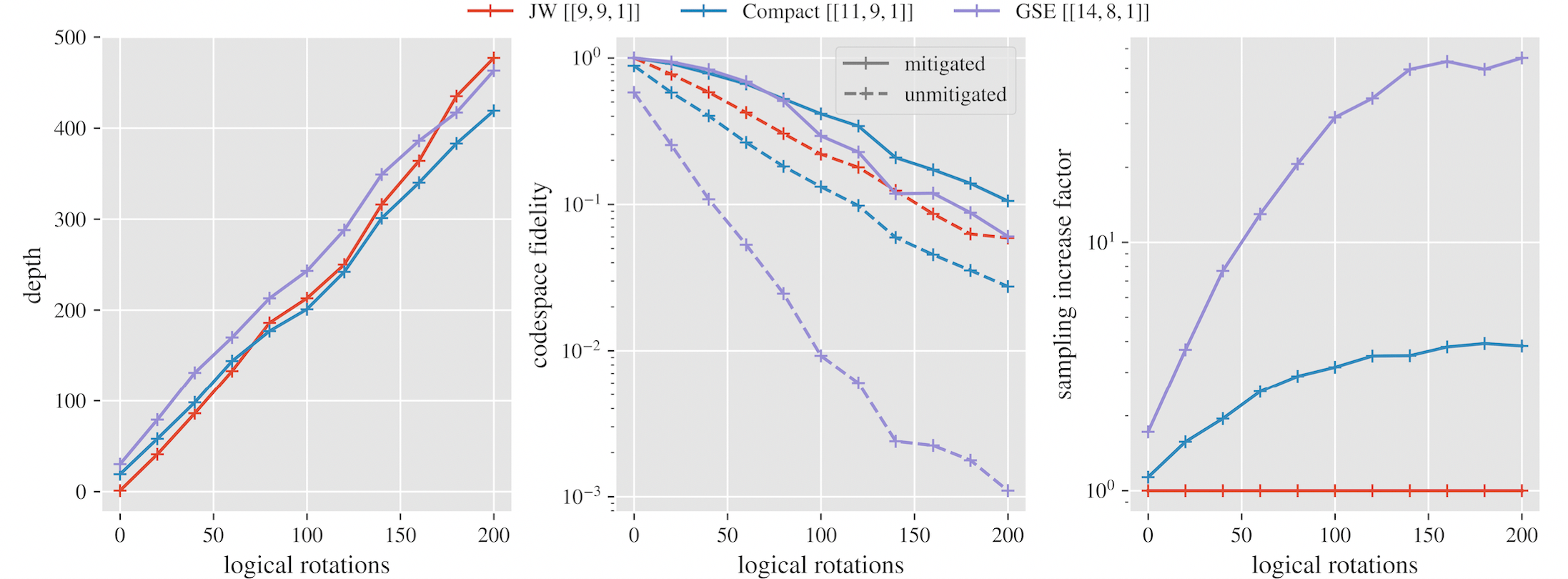}
\caption{Simulation results for the 9 mode random circuit simulations. For each number of rotations, the results are averaged over 100 different realizations of the randomly constructed circuits.}
\label{subfig:9_mode_random_circuits}
\end{subfigure}

\begin{subfigure}{\textwidth}
\includegraphics[width=\textwidth]{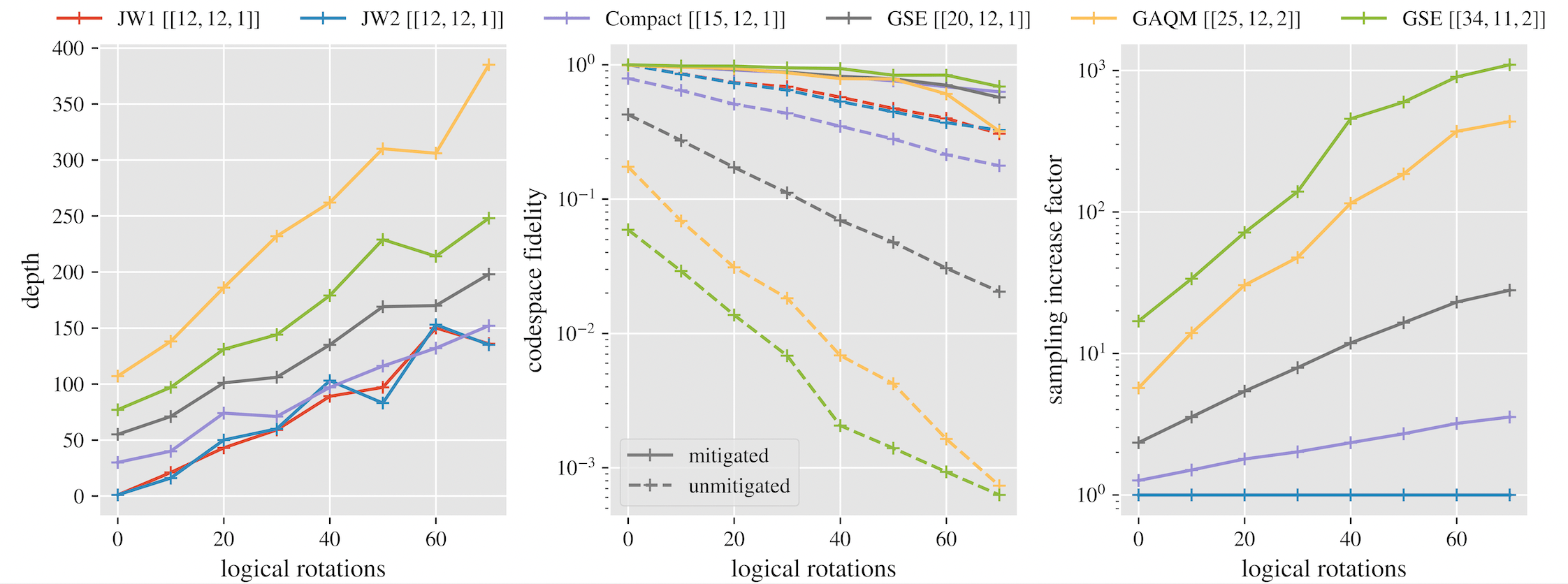}
\caption{Simulation results for the 12 mode random circuit simulations. For each number of rotations, the results are averaged over 10 different realizations of the randomly constructed circuits.}
\label{subfig:12_mode_random_circuits}
\end{subfigure}

\caption{(a) 9 mode random circuit experiments. For each number of logical rotations, the results are averaged over 100 randomly generated circuits. (b) 12 mode random circuit experiments. For each number of logical rotations, the results are averaged over 10 randomly generated circuits. `JW1' and `JW2' are two different linear orderings of the modes within the Jordan-Wigner transform. JW1 is a snake pattern across the left $2\times 3$ block attached to a snake the right $2\times 3$ block. JW2 is a horizontal snake pattern across the $4\times 3$ lattice.}
\label{fig:random_circuits}
\end{figure}

The results for the 9 mode experiments are shown in Fig. \ref{subfig:9_mode_random_circuits}. 

\subsection{Variational quantum eigensolver}

One fundamental task of quantum simulation is to approximate the ground state of  physical system Hamiltonians. A system of particular interest is the two-dimensional Fermi-Hubbard model due to  its relevance to studies of strongly-correlated electrons and high-temperature superconductivity. The Fermi-Hubbard Hamiltonian encourages fermions hopping between adjacent lattice sites $\langle i,j\rangle$ but biases fermions of opposite spin $\sigma$ being present at the same site:
\begin{equation}
    H_f = t\sum_{\mathclap{\substack{\langle i,j\rangle\\ \sigma \in \{\uparrow,\downarrow\}}}} a_{i,\sigma}^{\dag}a_{j,\sigma} + h.c. + U\sum_{i} n_{i,\uparrow}n_{i,\downarrow}\, .
\end{equation}
We will consider the model on a $2\times 3$ grid with $12$ fermionic modes away from half filling, where  spin-up and spin-down 2 particles  make for a filling fraction of $1/3$.

For the VQE circuit, we will use the Hamiltonian variational ansatz, which consists of a chosen number of Trotter steps evolving the system with rotations of Pauli strings from the model Hamiltonian, but the respective rotation angles are variational parameters set by gradient-descent optimization.
Specifically, a single iteration shall include each term of the Hamiltonian, except the trivial ones. The ansatz circuit has the form
\begin{equation}
    U(\vec{\theta}) = \prod_{j,\sigma} e^{  i\theta_j^{(1)} {\mathcal{B}}_{j,\sigma}}  \prod_j e^{ i\theta^{(2)}_j {\mathcal{B}}_{j,\uparrow}{\mathcal{B}}_{j,\downarrow}} \prod_{\langle j,k\rangle, \sigma} e^{ i\theta^{(3)}_{jk} {\mathcal{A}}_{jk,\sigma}{\mathcal{B}}_{k,\sigma} } e^{ i\theta^{(3)}_{jk} {\mathcal{B}}_{j,\sigma}{\mathcal{A}}_{jk,\sigma}}
\end{equation}
where $\vec{\theta}$ is the vector of all 32 angles $\theta^{(1)}_j$, $\theta^{(2)}_j$, $\theta^{(3)}_{jk}$ constituting the variational parameters. The two rotations appearing in the third product have the same angle $\theta^{(3)}_{jk}$ which ensures the conservation of the particle number. Our ansatz shall only consist of a single layer $U(\vec{\theta})$. The starting state is a Slater determinant state with the first two modes of each spin sector occupied.

The encodings used are presented in Table (\ref{tab:simulation_encodings}). The noise model is, as stated earlier,  1\% single-qubit depolarizing noise following every CNOT gate. 

Recall that when we calculate the expectation value of the Hamiltonian at the end of the simulation, we are doing so according to the equation
\begin{align}
   \langle H \rangle \; = \; \sum_p \alpha(p) \sum_{\lambda}\mathrm{tr}\left( \rho^{\phantom{\dagger}}_\lambda\Lambda^{p,\,\lambda}\right) \, .
\end{align}
where $\lambda$ is labeling the syndrome subspaces, $\alpha(p)$ is the coefficient for Pauli $p$ in the Pauli expansion of the Hamiltonian, $\Lambda^{p,\lambda}$ is the tapered Pauli operator $p$ in the syndrome subspace $\lambda$, and $\rho_{\lambda}$ is the density matrix component in the $\lambda$ subspace. 

\begin{figure}
    \centering
    \includegraphics[width=\linewidth]{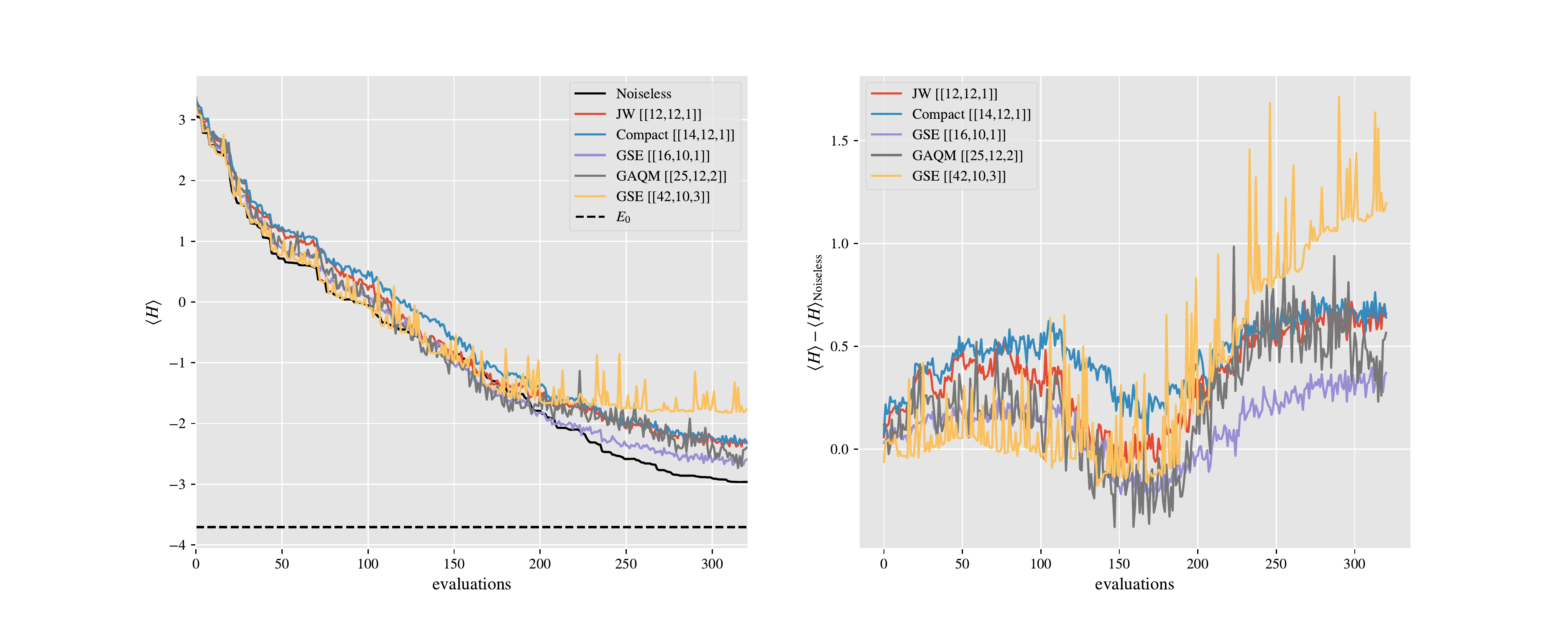}
    \caption{(left) Optimization curves for $2\times 3$-site (12 mode) Hubbard model with the shown mappings. On the x-axis is the number of times a parameter is updated in the one-gate-at-a-time update strategy we use. Each gate is updated 10 times throughout the optimization. (right) Difference between noisy error-mitigated optimization and the noiseless simulation}
    \label{fig:VQE}
\end{figure}

As the density matrix is reconstructed by sampling over noise realizations, we perform 1000 shots for each circuit that we simulate. Recall that this averaging over many shots is of a different nature than what is required in the averaging over shots that is required in a quantum experiment. Being only one layer, we do not expect our ansatz to be able to well approximate the ground state of the Hubbard Hamiltonian. Thus, instead of comparing how well each mapping is able to approximate the ground state during optimization, we will compare only the relative ability of the mappings to be optimized towards the ground state during the experiments when we simulate the error mitigation as compared to the noiseless simulation. 

In order to reduce the computation burden, we use a one-gate-at-a-time strategy to optimize the variational ansatz. We calculate a single gradient component at a time using the parameter shift rule from \cite{wierichs2022general} for each variational parameter and update the individual corresponding parameter instead of calculating the entire gradient vector and updating all the parameters at once. We then iterate over optimizing each parameter 10 times and compare the optimization curves across mappings. Each mapping begins with the same initial state and same variational parameters which are randomly chosen.

The results of the simulations are shown in Fig. \ref{fig:VQE}. We see that while the GSE $[[42,20,3]]$ and GAQM $[[25,12,2]]$ encodings offer greater ability to mitigate errors through postselection at the end of the computation, the result is that for a fixed shot budget, the unlikeliness with which the state is observed in the codespace adds a large degree of variance to the estimated expectation values possibly hampering the ability to optimize. 

\subsection{Non-equilibrium dynamics}

A basic primitive of quantum simulation is time evolution. One seeks to accurately approximate the unitary generated by a Hamiltonian by various decompositions, often (although not necessarily optimally) the Suzuki-Trotter decomposition. Accurate simulations for physically meaningful times however can lead to deep circuits and except in simple cases are often considered out of the reach of near-term quantum simulations. 

\begin{figure}
    \centering
    \includegraphics[width = \linewidth]{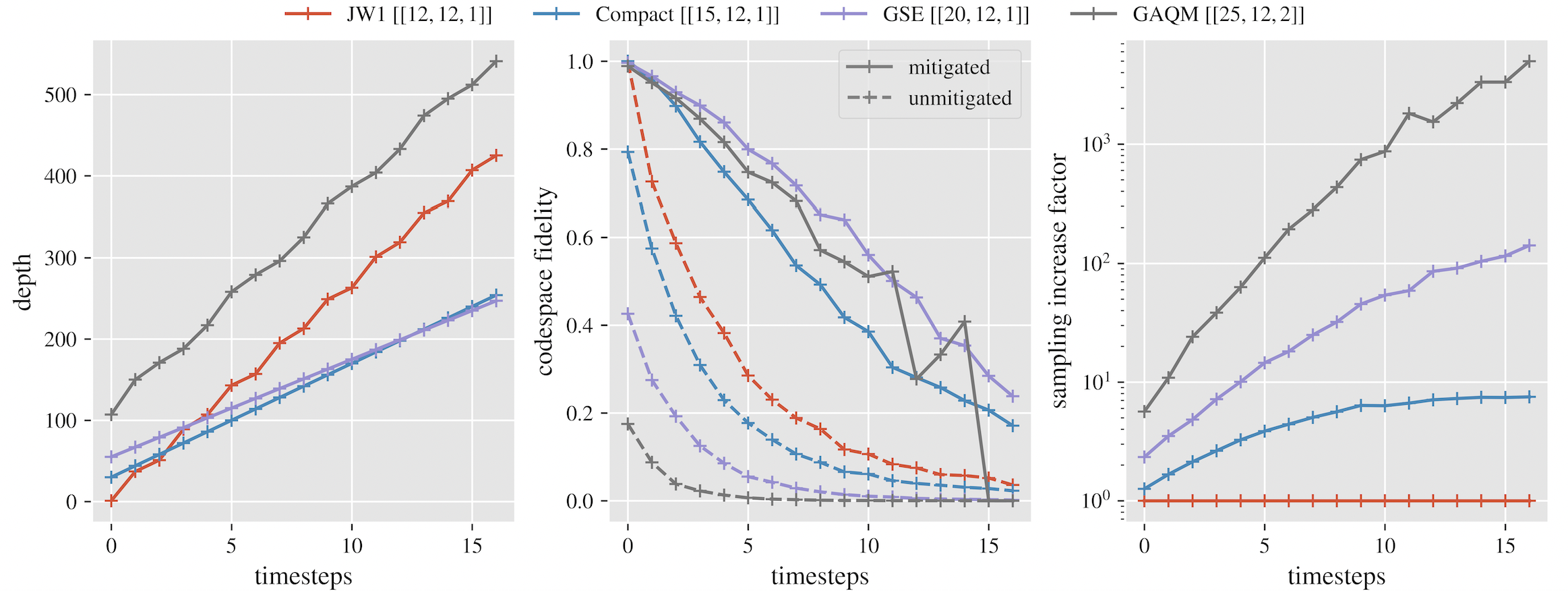}
    \caption{Simulation results for the simulations of the Rudner et. al. model on a $4\times 3$ lattice. \textbf{Left:} Depths for the circuits for the shown number of timesteps. \textbf{Center:} Fidelity of the noisy state against the ideal noiseless state for error mitigated and unmitigated simulations is shown. Note that this is shown on a linear scale. \textbf{Right:} The factor by which the number of taken samples must be increased by in order to have the same number of samples as without error mitigation.}
    \label{fig:floquet_circ}
\end{figure}

As a demonstration of our error mitigation scheme to a simple model of dynamics, we will consider a model proposed by Rudner et{.} al{.} \cite{rudner2013anomalous} in the study of edge modes in periodically driven systems. We choose this model in order to determine whether the error mitigation scheme we investigate here is useful in the observation of physically interesting phenomena in a digital quantum computer.
Rudner et. al.'s  model is a periodic unitary circuit where each period consists of a specially chosen sequence of four layers of unitary gates generated by hopping operators. The circuit for each period is given by
\begin{equation}
    U_T = \prod_{\langle i,j\rangle\in D} e^{\frac{i}{5}(a_i^{\dag}a_j + h.c.)} \prod_{\langle i,j\rangle\in C} e^{\frac{i}{5}(a_i^{\dag}a_j + h.c.)} \prod_{\langle i,j\rangle\in B} e^{\frac{i}{5}(a_i^{\dag}a_j + h.c.)} \prod_{\langle i,j\rangle\in A} e^{\frac{i}{5}(a_i^{\dag}a_j + h.c.)}
\end{equation}
The model is devised such that particles which start on the boundary of the system propagate unidirectionally around the system while particles in the interior of the system remain confined to small stationary orbits. In the original model, a fifth layer is included in each period consisting of exponentiated number operators that perturb the model away from the fixed point. We will opt to not include this perturbation so in the fixed point that we investigate, particles deterministically hop between lattice sites. Our focus is to study the extent to which the physics of the fine-tuned model could be recreated in an error mitigated computation on a noisy quantum computer.
We simulate the system on a $4\times 3$ lattice with four different mappings indicated in Fig. \ref{tab:simulation_encodings} with a number of qubits ranging from a 12 qubit Jordan-Wigner to a 25 qubit generalized AQM. The depths, fidelities, and sampling increase factors are shown in Fig. \ref{fig:floquet_circ}. The 15-qubit Compact encoding and the 20-qubit GSE provide much more efficient circuits than Jordan-Wigner. In the case of GSE, this can be partially attributed to the fact that the circuits do not include individual parity rotations for which JW is more efficient than GSE. At the $4\times 3$ size, GAQM does not at this size give any benefit over JW in terms of circuit depth due to the higher weight operators. After error mitigation, the GSE maintains the highest fidelity. This is attributed to the similar efficiency of the circuits as compared to the Compact encoding but the larger number of stabilizer generators (8 vs. 3) and the greater amount of postselection. By the end of the 16 timesteps, the factor by which the number of samples increases is over 100 and still increasing whereas the Compact encoding has seemingly saturated its diffusion out of the codespace. Assuming the state is maximally mixed across syndrome subspaces, the Compact encoding will require a factor of 8 increase in samples as there are 3 stabilizer generators. Finally, note that the fidelity for the GAQM drops precipitously after a certain point. This is due to the small probability of the final state begin measured in the codespace making it difficult to reconstruct an accurate density matrix without a prohibitively large number of samples.

\begin{figure}[t]
\centering
\begin{subfigure}[t]{0.36\textwidth}
\centering
\includegraphics[width = 0.95\linewidth]{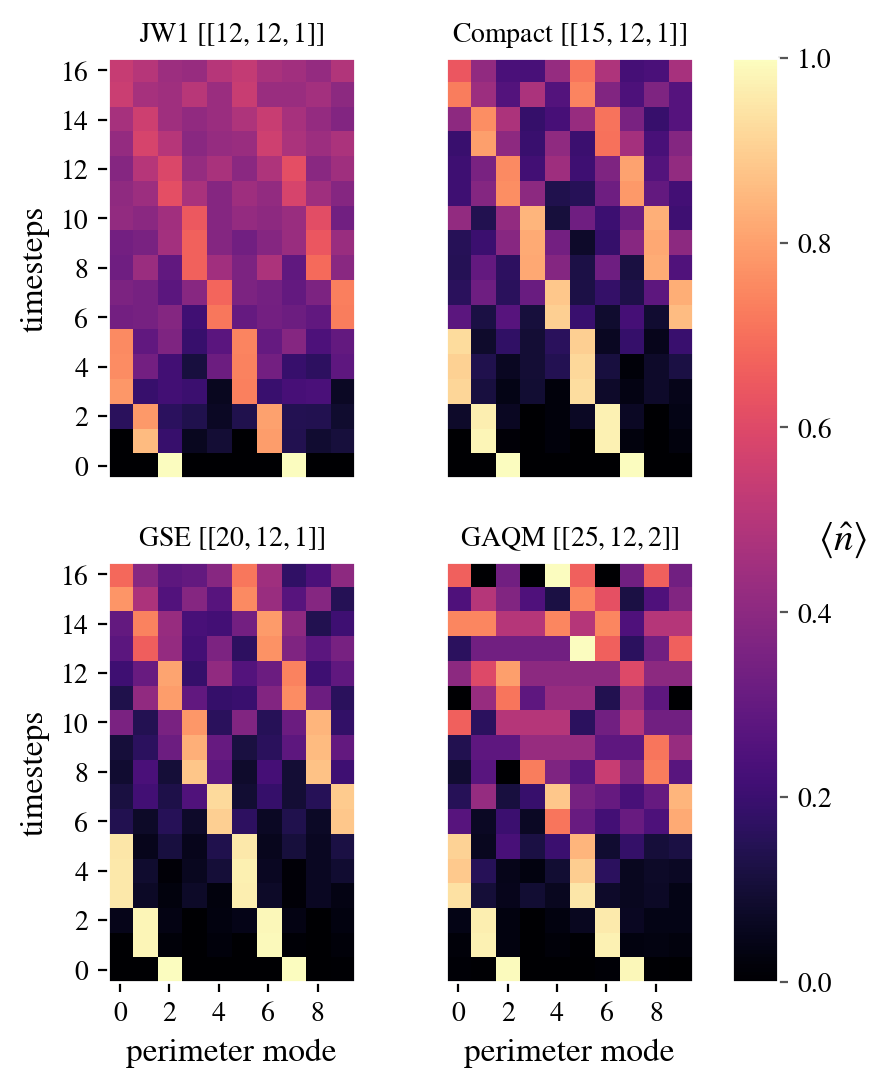}
\caption{}
\label{subfig:perimeter_mode_numbers}
\end{subfigure}
\begin{subfigure}[t]{0.62\textwidth}
\centering
\includegraphics[width = \linewidth]{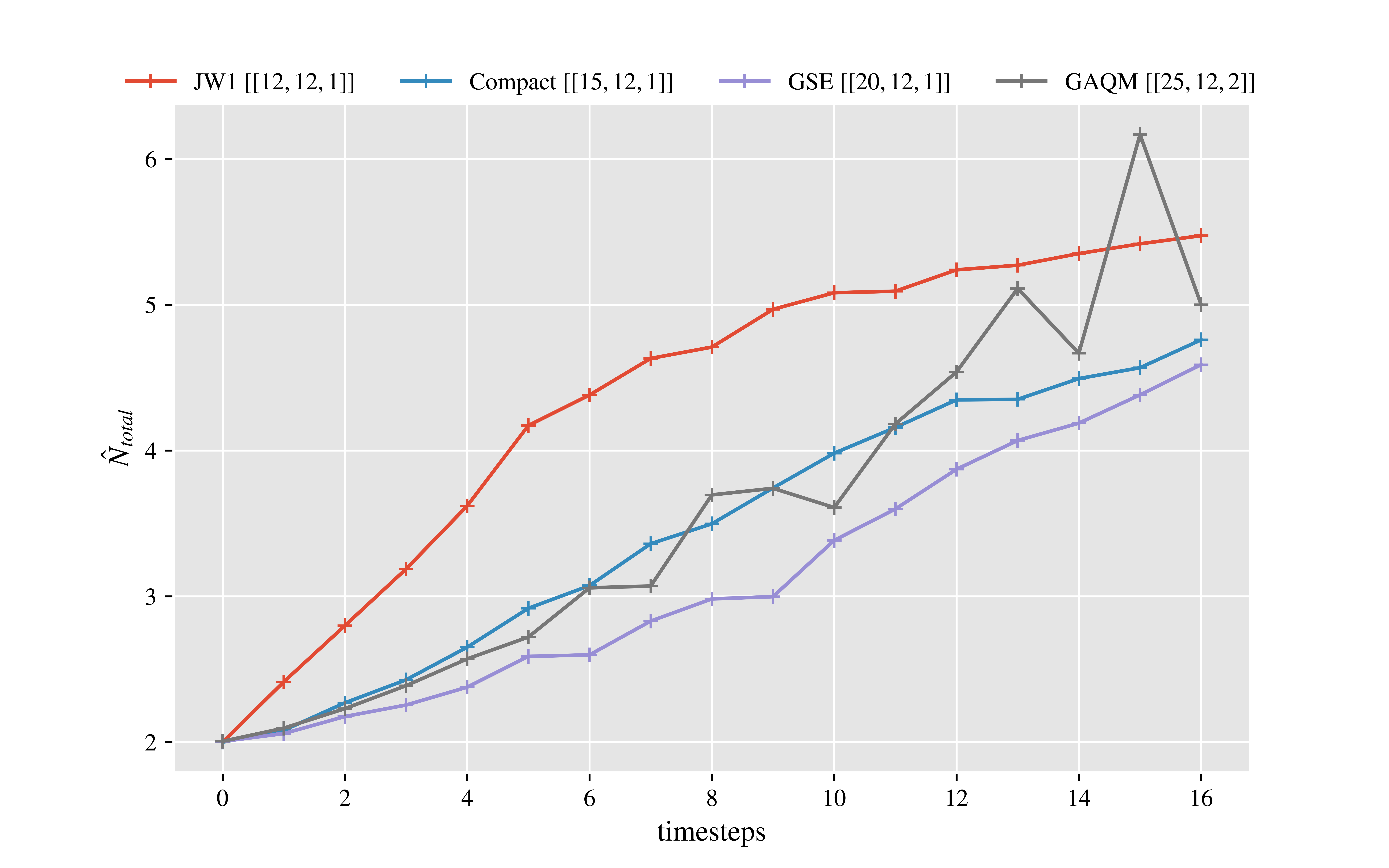}
\caption{}
\label{subfig:floquet_total_numbers}
\end{subfigure}

\caption{
(a) Expectation values of number operators for the 10 modes around the perimeter of the $4\times 3$ lattice for each mapping using stabilizer postselection. Not shown are the expectation values of the two interior modes. (b) The average total number of particles. In an ideal simulation, this would remain constant at 2.
}
\label{fig:floquet_numbers}
\end{figure}

We also track the evolution of observables throughout the dynamics. In particular, we investigate the extent to which the propagating edge modes can be observed in the noisy simulation. We initialize the system into a state with two particles occupying modes on opposite edges of the lattice. We then apply the encoded rotations as described above and at each timestep, calculate the expectation value of the number operators of the modes on the perimeter. In Fig. (\ref{subfig:perimeter_mode_numbers}), we have unrolled the 10 modes on the perimeter of the $4\times 3$ system and plotted the expectation values of the number operators. A noiseless simulation would show the particles propagating leftward as time goes upward on the plot. Every two or three timesteps, the particles deterministically hop to the next site. In our noisy simulations, we observe the number expectation values following the predicted pattern until the effects of noise begin to wash out the effect. In the case of the GAQM the probability of a given shot ending up in the codespace was small enough that only a few shots are being used to reconstruct the codespace density matrix. 

We also calculate the expectation value of the total number operator throughout the evolution. Because all rotations preserve total particle number, a noiseless simulation would have $\langle \hat{N}\rangle=2$ particles throughout the evolution. However, the effective action of the Pauli noise operators inside the codespace do not necessarily respect the total particle number symmetry. As the circuits get longer, noise changes the observed particle number in the direction of the maximally mixed value of half-filling. 
We observe that for all three nontrivial encodings, the observed total particle numbers for the remain closer to their initial value than the JW. For the Compact and GSE, this is due in part to the shorter circuits than JW (after the first few time steps). From the first few timesteps of GAQM, we can infer that the improvement is due in part as well to the codespace postselection as the circuits are for all timesteps shorter than JW.  

\section{Discussion}

In this work, we investigated encodings of fermionic systems equipped with a stabilizer group that provides a straightforward method of mitigating errors, through stabilizer postselection. Importantly, the stabilizer group arises naturally in the pursuit of methods to reduce circuit depths. They are a byproduct of seeking an encoding that maintains locality structures present in the target fermionic Hamiltonian, as doing so requires the introduction of additional qubits and the restriction to a subspace.

Focusing on two-dimensional quantum systems with local interactions, we performed numerical experiments simulating three distinct types of quantum simulation experiments (i) random, nearest-neighbor quantum circuits of fermionic unitaries on 2D lattices (ii) VQE for the Fermi-Hubbard model on a $2\times 3$ cluster (iii) discrete-time Floquet dynamics in a finely-tuned model of a floquet topological insulator where we investigated the visibility of chiral edge modes across multiple periods . 

In order to numerically investigate the stabilizer postselection error mitigation strategy, we introduced algorithms for performing classical simulations of quantum computations consisting of a sequence of logical operations on a stabilizer code, that are implemented through non-transversal operations. These logical operations have the form of exponentiated logical operators on the code. The presented simulation algorithms scale polynomially in the dimension of the code space (exponential in the number of encoded degrees of freedom) and polynomially in the number of stabilizer generators. The algorithms take advantage of the stabilizer formalism to allow for simulations of quantum circuits with Pauli noise acting on the physical qubits of the device.

The sizes of quantum experiments we considered lie near the boundary of the current state of the art for NISQ experiments and should help to inform the first experimental simulations of fermionic systems in 2D. All of our experiments featured 1\% single-qubit depolarizing noise channels that act after each two-qubit gate independently in parallel on the two involved qubits. Our random evolution and non-equilibrium dynamics numerical experiments show that already at the sizes of experiments we considered, there is an advantage in terms of the simulation fidelity and accuracy of observables to using a locality-preserving encoding over standard Jordan-Wigner when combined with stabilizer postselection. This increased accuracy comes at the cost of number of discarded shots that grows with the number of auxiliary qubits used and the depth of the circuit. For resource-frugal encodings such as the Derby-Klassen compact encoding which uses only a few additional qubits for the systems we investigated, the additional number of shots required is modest, less than a factor 10 increase. Note that the advantage gained is in spite of the overhead incurred by needing to initialize into the logical subspace. Without error mitigation, it seems to still be advantageous to use Jordan-Wigner at these system sizes.

Our simulations showed that it is important to prioritize low-weight encoded operators over high code distance in order to minimize circuit depth. Low-weight encoded operators, which have fewer physical qubits participating in the operation, allow for shallower circuits with fewer opportunities for errors to occur. While high code distance in principle allows one to detect more errors, our simulations demonstrated empirically that the benefits of low-weight operators in reducing circuit depth outweigh the potential benefits of higher code distance. However, such trade-offs may change as a function of the noise rate.

Our results have implications for other stabilizer-based error mitigation schemes as well. The stabilizer postselection scheme we considered requires making coherent measurements of the stabilizer generators which incurs additional errors as the stabilizer generators are measured. Other schemes for stabilizer-based error mitigation, such as subspace expansions, do not require making such coherent measurements of multi-qubit observables. Instead the projection is achieved at the level of expectation values by averaging over a larger set of observables. Idealized executions of both the stabilizer postselection (without noise in the stabilizer generator measurement modules) and subspace expansion in the large-sample limits would achieve the same expectation values for observables as both methods rely on making the same projection onto the codespace. In that regard, our results can provide imprecise and qualitative but potentially useful information about the potential performance of other symmetry-based error mitigation schemes. It will be important, however, for future work to make a quantitative comparison of other such error mitigation methods.

Additionally, as real experiments suffer from more general noise processes, it would be useful to investigate the effect of other types of noise that were not accessible to our numerical simulations. Finally, our results show that we should begin to consider quantum simulation of systems of lattice fermions as being within reach of the capabilities of modern devices.

\subsection*{Acknowledgement}
RWC and JDW were funded by the NSF (PHYS- 1820747) and the Department of Energy (Grant DE- SC0019374). JDW is also supported by NSF (EPSCoR- 1921199) and by the U.S. Department of Energy, Office of Science, Office of Advanced Scientific Computing Research under programs Quantum Computing Application Teams and Accelerated Research for Quantum Computing program.
\appendix
\section{Number of Hamiltonian terms}
\label{sec:n_terms}
As long as it acts on the same particle sectors, the exact number of Pauli strings in a Hamiltonian as well as its $\ell_1$ norm should always be constant over the use of different fermionic encodings. However, the lowest norm and number of terms might  be difficult to attain when we encode the Hamiltonian using quantum codes \cite{chien2019analysis}. Let us sketch the cause for this issue before explaining the fix. In this investigation we find that mappings like \eqref{eq:mapping} often fail to capture cancellations of Hamiltonian terms from double excitation operators. Operators $a^{\dagger}_ja^{\phantom{\dagger}}_ka^{\dagger}_la^{\phantom{\dagger}}_m$ from four distinct modes $(j,k,l,m)$ that occur in a Hamiltonian together with operators in which  $(j,k,l,m)$ permuted often result in Pauli strings that should cancel or combine, but do not as they differ by the multiplication of stabilizers such as
    \begin{align}
        \mathcal{A}_{jk}\mathcal{A}_{kl}\mathcal{A}_{lm}\mathcal{A}_{mj}\, .
    \end{align}
    This does not mean the mapping is wrong, as logical operators may differ by multiplication of stabilizers in general, but the mapping causes a proliferation of terms in the Hamiltonian. To cure a Hamiltonian suffering from such a proliferation, we have developed a classical routine with polynomial runtime in the number of Hamiltonian terms and the number of stabilizer generators. This is exponentially better than the na\"ive approach of comparing every pair of terms under multiplication of the entire stabilizer group. Let us sketch the algorithm: we start by applying \textbf{Algorithm$\,$1} to a Hamiltonian like in \eqref{eq:reduce}, but we keep track of which  Pauli strings $p$  transform into what $\Lambda^p$. The reason for this is that there might be some two Pauli strings $p^{\phantom{\prime}},p^\prime$ such that $\Lambda^p=\pm \Lambda^{p^\prime}$. We watch them recombine and cancel as 
    \begin{align}
    \label{eq:cure1}
    \sum_p \alpha(p) \, p^{\phantom{p}}_{[1,n]} \; &\mapsto \; \sum_p \alpha(p) \, \Lambda^{p}_{[1,n-r]} \\ &= \; \sum_{q} \left( \sum_{(p,s): \, \Lambda^p = (-1)^s q} \left( -1\right)^s\,  {\alpha}(p)\right) q_{[1,n-r]}^{\phantom{p}} \, . \label{eq:cure2}
    \end{align}
Note that  $p$ and $q$ are unsigned Pauli strings, whereas $\Lambda^p$ is by definition signed and $s\in \lbrace 0,1\rbrace$. For every $q_{[1,n-r]}$ we select exactly one $(-1)^sp_{[1,n]}$ from the record such that $\Lambda^{p} = (-1)^s q$. We now replace  $q_{[1,n-r]}$ in \eqref{eq:cure2}  with that operator. There could be some added benefit in choosing to replace $q$ with one $(-1)^s p$ that $p$ has the lowest weight or the best locality amongst all operators on the record of $q$. The Hamiltonian has subsequently received a consistent representation of its logical operators, which has minimized its number of terms and $\ell_1$ norm.\\

\section{State preparation}
\label{sec:stateprep}
Due to our use of a quantum code, we need to prepare an initial state that is in the $+1$ eigenspace of each stabilizer generator $\lbrace S_k \rbrace$. We also need to prepare  a deterministic logical state in the right particle number manifold: this state can easily be the stabilizer state of some set logical operators. A sensible choice would be  (a subset of) the vertex operators $\lbrace\pm \mathcal{B}_l\rbrace$, modified by signs to fix the correct particle number. The preparation of the code space and the logical state can then be combined. Extending the list of stabilizers by $n-r$ logical operators, the system can be constrained completely using projective measurements of $\lbrace S_k \rbrace$ and $\lbrace\pm \mathcal{B}_l\rbrace$. Let us describe how this state preparation will look like on a quantum computer, and then let us see how it is influenced by Pauli noise, before we turn to discuss how the noisy state preparation is handled in the simulator.\\
A single Pauli string can be measured by a quantum circuit similar to the Pauli string rotation in Figure \ref{fig:circuit}(a), but with a $Z$-measurement in place of the rotation $R_z$. While such a measurement fixes the state, we have no guarantee that the measurement has the correct outcome. We could of course interpret opposite outcomes as syndromes and then attempt quantum error correction, but we would much rather re-define the quantum code on the fly such that we are in the correct subspace. For auxiliary qubit codes, for instance, we are free to attach minus signs to vertex operators and stabilizers after the fact. In superfast codes we cannot attach signs to stabilizers directly, so we must attach them to exchange operators over graph edges $\widehat{\mathcal{A}}_{jk}$. For some of these codes, such as the Generalized Superfast encoding, the minus signs on edge operators fix the particle parity sector, as not all vertex operators are nontrivial. When attaching minus signs to stabilizers and logical operators in that way, ansatz operators and observables need to be recompiled after the measurement outcomes are known, but it renders the state preparation deterministic.\\
The Clifford circuits around the measurements can of course be noisy, and so we cannot completely trust the measurement outcomes. After the state is projected into a subspace, it can be flipped into a syndrome space by Pauli noise. We could have measured a syndrome pattern $\nu \in \lbrace 0, 1\rbrace^n$, but in reality, we have a  mixture of states with  all kinds of syndrome patterns  $\mu \in \lbrace 0, 1\rbrace^n$. One could describe this classical-quantum state by
\begin{align}
    \label{eq:quantumclassical}
    \sum_{\mu, \nu \in\lbrace 0, 1\rbrace^n} 
    \widehat{\rho}^{\phantom{\dagger}}_{\mu, \nu} \, \left| \mu \right\rangle\!\!\left\langle \mu \right|^{\phantom{\dagger}}_{[1, n]} \otimes \left|  \nu \right\rangle\!\!\left\langle \nu \right|^{\phantom{\dagger}}_{(n, 2n]} \, ,
\end{align}
where qubits $(n, 2n]$ denote the flags of the measurement outcomes as they are recorded, and $\widehat{\rho}_{\mu, \nu}$ are positive numbers describing the quantum-classical distribution between recorded measurements $\nu$ and actual state syndromes $\mu$. Following the measurement of a pattern $\nu$, we can run a time evolution / ansatz circuit adapted to it, and so we only care about the deviations of the actual syndromes from the recorded syndromes $\mu \oplus \nu$, rather than $\mu$ and $\nu$ themselves. We can therefore average all classical branches in \eqref{eq:quantumclassical} by summing them pretending that  all $\nu$ are $ 0^{ n} $ and all $\mu$ are $\mu \oplus \nu$. The resulting density matrix can be related to \eqref{eq:blocks} when taking into account that the last $n-r$ bits of the syndromes correspond to measurement outcomes of logical operators. Let  $M_1,\, M_2, \, \dots ,\, M_{n-r}$ denote these logical operators in their physical representation. We can obtain the syndrome blocks of \eqref{eq:blocks} from \eqref{eq:quantumclassical} by

\begin{align}
\label{eq:initial_state}
     \rho^{\phantom{\dagger}}_\lambda = \sum_{\mu \in\lbrace 0, 1\rbrace^n} \sum_{\omega \in\lbrace 0, 1\rbrace^{n-r}} \widehat{\rho}^{\phantom{\dagger}}_{\mu, \,\mu \oplus (\lambda|\omega)} \left[\prod^{n-r}_{k=1}\frac{1}{\sqrt{2}} \left( \vphantom{\frac{\sum}{\sum}}\id  + (-1)^{\mu_{r + k} + \omega_k }\Lambda^{M_k}\right)\right]|\mathcal{R}\rangle\!\langle \mathcal{R}| \left[\prod^{n-r}_{l=1}\frac{1}{\sqrt{2}} \left( \vphantom{\frac{\sum}{\sum}}\id  + (-1)^{\mu_{r + l} + \omega_l }\Lambda^{M_l}\right)\right] \, 
\end{align}
where $(\lambda|\omega) = (\lambda_1,\, \dots,\, \lambda_r, \omega_1,\, \dots , \, \omega_{n-r})$ and where $|\mathcal{R}\rangle$ is a suitable product state similar to $\bigotimes_l |w^{(l)}\rangle_l$ in \eqref{eq:stabstate}, but with respect to  $\lbrace\Lambda^{M_l}\rbrace$  as stabilizers.

In the \simulator simulator, every shot of the entire circuit would start out as a shot of \eqref{eq:initial_state}, where a syndrome pattern  with respect to stabilizer generators and logical operators is developed. As the  state at the end of the state preparation is an $n$-qubit stabilizer state, we can simulate it as a list of $n$ syndromes only during the state preparation: statistically planted errors are propagated through the Clifford part of the measurement circuit just like in \textbf{Algorithm$\,$2}, but we only need to check the commutation relations of the resulting Pauli strings with the $n$ elements of the extended stabilizer list to decide which bits must be flipped in the syndrome list. The center of every measurement subcircuit however will reset one of the bits. Errors, that might have flipped that bit before the measurement are irrelevant, as its value in the code space is fixed at the point of measurement. However, we would leave the code space if the bit is flipped later. During the simulation of the state preparation, the conversion of physical to logical operators is not necessary. The conversion takes place only after the routine, when the last $n-r$ bits of the syndromes $(\lambda|s)$ are used to create a logical stabilizer state
\begin{align}
    \left[\prod^{n-r}_{k=1}\frac{1}{\sqrt{2}} \left( \vphantom{\frac{\sum}{\sum}}\id  + \left(-1\right)^{s_k} \Lambda^{M_k}\right)\right]|\mathcal{R}\rangle
\end{align}\, 
and the first $r$ bits decide its syndromes $\lambda$.

\section{Generalized auxiliary qubit code}
\label{sec:genaux}
The auxiliary qubit code \cite{steudtner2019quantum} is a $[[n,N,1]]$ quantum code defined on top of a $N$-qubit Jordan-Wigner transform under the addition of $n-N$ auxiliary qubits. The auxiliary qubit code allows us to detect single Pauli errors on the $N$ original qubits, but is blind against bit flip errors on the qubits that were added. We here present a generalized version of the auxiliary qubit code with a code distance of two. Let us start with its construction. We define an invertible binary $(n-N)\times(n-N)$ matrix $M$ as a classical encoding layer on the auxiliary qubits (labeled $N+1, \dots, n$). This classical encoding layer turns Pauli operators $Z_{N+j}\mapsto$, $X_{N+k}$ into Pauli strings
\begin{align}
    {\mathfrak{Z}^j_{(N, n]}} \;=& \bigotimes_{k:\, M^{-1}_{jk}=1 }  Z^{\phantom{\dagger}}_{N+k} \, ,\\
    {\mathfrak{X}^k_{(N, n]}} \;=& \bigotimes_{j:\,  M_{jk}=1 }  X^{\phantom{\dagger}}_{N+k}
\end{align}
such that $\mathfrak{Z}^j$ and $\mathfrak{X}^k$  anticommute when $j=k$. A set of  stabilizer generators is then defined by $n-N$ signed Pauli strings $t^k$ on the first $N$ qubits. In contrast to the original auxiliary qubit mapping, these strings do not necessarily need to commute -- in fact we want a few of them to anticommute such that we can detect errors on the auxiliary qubits. Let us denote the anticommutator of two operators $a$, $b$ by $\lbrace a,b\rbrace = ab + ba$. A set of stabilizer generators $Q^{k}$ is then defined as
\begin{align}
    Q^{k}_{[1,n]} \;=\; t^k_{[1,N]}\otimes {\mathfrak{X}^k_{(N, n]}} \prod_{{\substack{j<k: \\ \lbrace t^j,\, t^k\rbrace=0}}} {\mathfrak{Z}^j_{(N, n]}} \, .
\end{align}
To comply with the stabilizer conditions, every Pauli string $p$ coming out of the Jordan-Wigner transform must be adjusted by
\begin{align}
\label{eq:auxlogical}
    p^{\phantom{\dagger}}_{[1,N]} \;\mapsto\; p^{\phantom{\dagger}}_{[1,N]} \prod_{m: \, \lbrace p,\,t^m\rbrace=0} {\mathfrak{Z}^m_{(N, n]}} \, .
\end{align}
The idea is now for the Pauli strings $t^k$ to be nonlocal, such that they can cancel nonlocal operators $p$. The vertex operators $\mathcal{B}_k$ and exchange operators $\mathcal{A}_{jk}$ of the Jordan-Wigner transform  for $j<k$  are
\begin{align}
    {\mathcal{A}_{{jk}_{[1,N]}}^{\text{(JW)}}} = Y^{\phantom{\dagger}}_{j}\otimes Z^{\phantom{\dagger}}_{j+1}  \otimes \cdots \otimes Z^{\phantom{\dagger}}_{k-1} \otimes X^{\phantom{\dagger}}_{k} \quad \text{and} \quad \mathcal{B}^{\, \text{(JW)}}_{k_{[1,N]}} = Z^{\phantom{\dagger}}_{k} \, .
\end{align}
where $\mathcal{A}_{jk}$ can be nonlocal.  Let there be an index $\hat{m}$ for which $t^{\hat{m}} = \mathcal{A}^{\text{(JW)}}_{jk}$, such that there is an exchange operator for the generalized auxiliary qubit code with 
\begin{align}
    {\mathcal{A}_{jk}}_{[1,n]} \; &= \; Q^{\hat{m}}_{[1,n]}  \,\cdot\, {\mathcal{A}_{{jk}_{[1,N]}}^{\text{(JW)}}} \prod_{m: \lbrace \mathcal{A}_{jk},\,t^m\rbrace=0} \mathfrak{Z}^{m}_{(N,n]} \\ \label{eq:auxexchange}
    &=  \; \mathfrak{X}^{\hat{m}}_{(N,n]} \prod_{\substack{l>\hat{m}: \\ \lbrace \mathcal{A}_{jk},\,t^l\rbrace=0}} \mathfrak{Z}^{l}_{(N,n]} \, ,
\end{align}
which can be made nonlocal. Defining all stabilizers $Q^m$ to encode Jordan-Wigner exchange operators along a connected path $t^m = \mathcal{A}^{\text{(JW)}}_{j_{m}j_{m+1}}$ would avoid nonlocal $\mathfrak{Z}$-products in logical operators  \eqref{eq:auxlogical}. It also allow us to catch a combination of bit flip errors equal to $\mathfrak{X}^{l}$ with stabilizer $Q^{l+1}$. We can almost set the binary matrix $M$ equal to the identity, which would mean $\mathfrak{Z}^k_{(N,n]} = Z_{N+k}$ and $\mathfrak{X}^k_{(N,n]} = X_{N+k}$ for all $k \in [1,n-N]$, but then we would be blind to bit flip errors on the last qubit $X_{n}$, as there is no stabilizer  $Q^{n+1}$ that could detect it. To tie off that last qubit, we require a small non-diagonal block in $M$. One can detect all single-qubit errors on the original $N$ qubits when the set of chain links $\lbrace (j_1,j_2), (j_2,j_3), \dots\rbrace$ of $t^m = \mathcal{A}^{\text{(JW)}}_{j_{m}j_{m+1}}$ features all  indices in $[1,N]$ at least twice. When all these conditions are met, the generalized auxiliary qubit code is at least distance two.

\bibliography{main.bbl}

\begin{thebibliography}{36}%
\makeatletter
\providecommand \@ifxundefined [1]{%
 \@ifx{#1\undefined}
}%
\providecommand \@ifnum [1]{%
 \ifnum #1\expandafter \@firstoftwo
 \else \expandafter \@secondoftwo
 \fi
}%
\providecommand \@ifx [1]{%
 \ifx #1\expandafter \@firstoftwo
 \else \expandafter \@secondoftwo
 \fi
}%
\providecommand \natexlab [1]{#1}%
\providecommand \enquote  [1]{``#1''}%
\providecommand \bibnamefont  [1]{#1}%
\providecommand \bibfnamefont [1]{#1}%
\providecommand \citenamefont [1]{#1}%
\providecommand \href@noop [0]{\@secondoftwo}%
\providecommand \href [0]{\begingroup \@sanitize@url \@href}%
\providecommand \@href[1]{\@@startlink{#1}\@@href}%
\providecommand \@@href[1]{\endgroup#1\@@endlink}%
\providecommand \@sanitize@url [0]{\catcode `\\12\catcode `\$12\catcode
  `\&12\catcode `\#12\catcode `\^12\catcode `\_12\catcode `\%12\relax}%
\providecommand \@@startlink[1]{}%
\providecommand \@@endlink[0]{}%
\providecommand \url  [0]{\begingroup\@sanitize@url \@url }%
\providecommand \@url [1]{\endgroup\@href {#1}{\urlprefix }}%
\providecommand \urlprefix  [0]{URL }%
\providecommand \Eprint [0]{\href }%
\providecommand \doibase [0]{https://doi.org/}%
\providecommand \selectlanguage [0]{\@gobble}%
\providecommand \bibinfo  [0]{\@secondoftwo}%
\providecommand \bibfield  [0]{\@secondoftwo}%
\providecommand \translation [1]{[#1]}%
\providecommand \BibitemOpen [0]{}%
\providecommand \bibitemStop [0]{}%
\providecommand \bibitemNoStop [0]{.\EOS\space}%
\providecommand \EOS [0]{\spacefactor3000\relax}%
\providecommand \BibitemShut  [1]{\csname bibitem#1\endcsname}%
\let\auto@bib@innerbib\@empty
\bibitem [{\citenamefont {Bravyi}\ and\ \citenamefont
  {Kitaev}(2002)}]{bravyi2002fermionic}%
  \BibitemOpen
  \bibfield  {author} {\bibinfo {author} {\bibfnamefont {S.~B.}\ \bibnamefont
  {Bravyi}}\ and\ \bibinfo {author} {\bibfnamefont {A.~Y.}\ \bibnamefont
  {Kitaev}},\ }\bibfield  {title} {\bibinfo {title} {Fermionic quantum
  computation},\ }\href@noop {} {\bibfield  {journal} {\bibinfo  {journal}
  {Annals of Physics}\ }\textbf {\bibinfo {volume} {298}},\ \bibinfo {pages}
  {210} (\bibinfo {year} {2002})}\BibitemShut {NoStop}%
\bibitem [{\citenamefont {Ball}(2005)}]{ball2005fermions}%
  \BibitemOpen
  \bibfield  {author} {\bibinfo {author} {\bibfnamefont {R.}~\bibnamefont
  {Ball}},\ }\bibfield  {title} {\bibinfo {title} {Fermions without fermion
  fields},\ }\href@noop {} {\bibfield  {journal} {\bibinfo  {journal} {Physical
  review letters}\ }\textbf {\bibinfo {volume} {95}},\ \bibinfo {pages}
  {176407} (\bibinfo {year} {2005})}\BibitemShut {NoStop}%
\bibitem [{\citenamefont {Verstraete}\ and\ \citenamefont
  {Cirac}(2005)}]{verstraete2005mapping}%
  \BibitemOpen
  \bibfield  {author} {\bibinfo {author} {\bibfnamefont {F.}~\bibnamefont
  {Verstraete}}\ and\ \bibinfo {author} {\bibfnamefont {J.~I.}\ \bibnamefont
  {Cirac}},\ }\bibfield  {title} {\bibinfo {title} {Mapping local hamiltonians
  of fermions to local hamiltonians of spins},\ }\href@noop {} {\bibfield
  {journal} {\bibinfo  {journal} {Journal of Statistical Mechanics: Theory and
  Experiment}\ }\textbf {\bibinfo {volume} {2005}},\ \bibinfo {pages} {P09012}
  (\bibinfo {year} {2005})}\BibitemShut {NoStop}%
\bibitem [{\citenamefont {Whitfield}\ \emph {et~al.}(2016)\citenamefont
  {Whitfield}, \citenamefont {Havl{\'\i}{\v{c}}ek},\ and\ \citenamefont
  {Troyer}}]{whitfield2016local}%
  \BibitemOpen
  \bibfield  {author} {\bibinfo {author} {\bibfnamefont {J.~D.}\ \bibnamefont
  {Whitfield}}, \bibinfo {author} {\bibfnamefont {V.}~\bibnamefont
  {Havl{\'\i}{\v{c}}ek}},\ and\ \bibinfo {author} {\bibfnamefont
  {M.}~\bibnamefont {Troyer}},\ }\bibfield  {title} {\bibinfo {title} {Local
  spin operators for fermion simulations},\ }\href@noop {} {\bibfield
  {journal} {\bibinfo  {journal} {Physical Review A}\ }\textbf {\bibinfo
  {volume} {94}},\ \bibinfo {pages} {030301} (\bibinfo {year}
  {2016})}\BibitemShut {NoStop}%
\bibitem [{\citenamefont {Chen}\ \emph {et~al.}(2018)\citenamefont {Chen},
  \citenamefont {Kapustin},\ and\ \citenamefont
  {Radi{\v{c}}evi{\'c}}}]{chen2018exact}%
  \BibitemOpen
  \bibfield  {author} {\bibinfo {author} {\bibfnamefont {Y.-A.}\ \bibnamefont
  {Chen}}, \bibinfo {author} {\bibfnamefont {A.}~\bibnamefont {Kapustin}},\
  and\ \bibinfo {author} {\bibfnamefont {{\DJ}.}~\bibnamefont
  {Radi{\v{c}}evi{\'c}}},\ }\bibfield  {title} {\bibinfo {title} {Exact
  bosonization in two spatial dimensions and a new class of lattice gauge
  theories},\ }\href@noop {} {\bibfield  {journal} {\bibinfo  {journal} {Annals
  of Physics}\ }\textbf {\bibinfo {volume} {393}},\ \bibinfo {pages} {234}
  (\bibinfo {year} {2018})}\BibitemShut {NoStop}%
\bibitem [{\citenamefont {Steudtner}\ and\ \citenamefont
  {Wehner}(2019)}]{steudtner2019quantum}%
  \BibitemOpen
  \bibfield  {author} {\bibinfo {author} {\bibfnamefont {M.}~\bibnamefont
  {Steudtner}}\ and\ \bibinfo {author} {\bibfnamefont {S.}~\bibnamefont
  {Wehner}},\ }\bibfield  {title} {\bibinfo {title} {Quantum codes for quantum
  simulation of fermions on a square lattice of qubits},\ }\href@noop {}
  {\bibfield  {journal} {\bibinfo  {journal} {Physical Review A}\ }\textbf
  {\bibinfo {volume} {99}},\ \bibinfo {pages} {022308} (\bibinfo {year}
  {2019})}\BibitemShut {NoStop}%
\bibitem [{\citenamefont {Chiew}\ and\ \citenamefont
  {Strelchuk}(2021)}]{chiew2021optimal}%
  \BibitemOpen
  \bibfield  {author} {\bibinfo {author} {\bibfnamefont {M.}~\bibnamefont
  {Chiew}}\ and\ \bibinfo {author} {\bibfnamefont {S.}~\bibnamefont
  {Strelchuk}},\ }\bibfield  {title} {\bibinfo {title} {Optimal fermion-qubit
  mappings},\ }\href@noop {} {\bibfield  {journal} {\bibinfo  {journal}
  {arXiv:2110.12792}\ } (\bibinfo {year} {2021})}\BibitemShut {NoStop}%
\bibitem [{\citenamefont {O'Brien}\ and\ \citenamefont
  {Strelchuk}(2022)}]{o2022ultrafast}%
  \BibitemOpen
  \bibfield  {author} {\bibinfo {author} {\bibfnamefont {O.}~\bibnamefont
  {O'Brien}}\ and\ \bibinfo {author} {\bibfnamefont {S.}~\bibnamefont
  {Strelchuk}},\ }\bibfield  {title} {\bibinfo {title} {Ultrafast hybrid
  fermion-to-qubit mapping},\ }\href@noop {} {\bibfield  {journal} {\bibinfo
  {journal} {arXiv:2211.16389}\ } (\bibinfo {year} {2022})}\BibitemShut
  {NoStop}%
\bibitem [{\citenamefont {Setia}\ \emph {et~al.}(2019)\citenamefont {Setia},
  \citenamefont {Bravyi}, \citenamefont {Mezzacapo},\ and\ \citenamefont
  {Whitfield}}]{setia2019superfast}%
  \BibitemOpen
  \bibfield  {author} {\bibinfo {author} {\bibfnamefont {K.}~\bibnamefont
  {Setia}}, \bibinfo {author} {\bibfnamefont {S.}~\bibnamefont {Bravyi}},
  \bibinfo {author} {\bibfnamefont {A.}~\bibnamefont {Mezzacapo}},\ and\
  \bibinfo {author} {\bibfnamefont {J.~D.}\ \bibnamefont {Whitfield}},\
  }\bibfield  {title} {\bibinfo {title} {Superfast encodings for fermionic
  quantum simulation},\ }\href@noop {} {\bibfield  {journal} {\bibinfo
  {journal} {Physical Review Research}\ }\textbf {\bibinfo {volume} {1}},\
  \bibinfo {pages} {033033} (\bibinfo {year} {2019})}\BibitemShut {NoStop}%
\bibitem [{\citenamefont {Jiang}\ \emph {et~al.}(2019)\citenamefont {Jiang},
  \citenamefont {McClean}, \citenamefont {Babbush},\ and\ \citenamefont
  {Neven}}]{jiang2019majorana}%
  \BibitemOpen
  \bibfield  {author} {\bibinfo {author} {\bibfnamefont {Z.}~\bibnamefont
  {Jiang}}, \bibinfo {author} {\bibfnamefont {J.}~\bibnamefont {McClean}},
  \bibinfo {author} {\bibfnamefont {R.}~\bibnamefont {Babbush}},\ and\ \bibinfo
  {author} {\bibfnamefont {H.}~\bibnamefont {Neven}},\ }\bibfield  {title}
  {\bibinfo {title} {Majorana loop stabilizer codes for error mitigation in
  fermionic quantum simulations},\ }\href@noop {} {\bibfield  {journal}
  {\bibinfo  {journal} {Physical Review Applied}\ }\textbf {\bibinfo {volume}
  {12}},\ \bibinfo {pages} {064041} (\bibinfo {year} {2019})}\BibitemShut
  {NoStop}%
\bibitem [{\citenamefont {Derby}\ \emph {et~al.}(2021)\citenamefont {Derby},
  \citenamefont {Klassen}, \citenamefont {Bausch},\ and\ \citenamefont
  {Cubitt}}]{derby2021compact}%
  \BibitemOpen
  \bibfield  {author} {\bibinfo {author} {\bibfnamefont {C.}~\bibnamefont
  {Derby}}, \bibinfo {author} {\bibfnamefont {J.}~\bibnamefont {Klassen}},
  \bibinfo {author} {\bibfnamefont {J.}~\bibnamefont {Bausch}},\ and\ \bibinfo
  {author} {\bibfnamefont {T.}~\bibnamefont {Cubitt}},\ }\bibfield  {title}
  {\bibinfo {title} {Compact fermion to qubit mappings},\ }\href@noop {}
  {\bibfield  {journal} {\bibinfo  {journal} {Physical Review B}\ }\textbf
  {\bibinfo {volume} {104}},\ \bibinfo {pages} {035118} (\bibinfo {year}
  {2021})}\BibitemShut {NoStop}%
\bibitem [{\citenamefont {Bausch}\ \emph {et~al.}(2020)\citenamefont {Bausch},
  \citenamefont {Cubitt}, \citenamefont {Derby},\ and\ \citenamefont
  {Klassen}}]{bausch2020mitigating}%
  \BibitemOpen
  \bibfield  {author} {\bibinfo {author} {\bibfnamefont {J.}~\bibnamefont
  {Bausch}}, \bibinfo {author} {\bibfnamefont {T.}~\bibnamefont {Cubitt}},
  \bibinfo {author} {\bibfnamefont {C.}~\bibnamefont {Derby}},\ and\ \bibinfo
  {author} {\bibfnamefont {J.}~\bibnamefont {Klassen}},\ }\bibfield  {title}
  {\bibinfo {title} {Mitigating errors in local fermionic encodings},\
  }\href@noop {} {\bibfield  {journal} {\bibinfo  {journal} {arXiv preprint
  arXiv:2003.07125}\ } (\bibinfo {year} {2020})}\BibitemShut {NoStop}%
\bibitem [{\citenamefont {Derby}\ and\ \citenamefont
  {Klassen}(2021)}]{derby2021compact2}%
  \BibitemOpen
  \bibfield  {author} {\bibinfo {author} {\bibfnamefont {C.}~\bibnamefont
  {Derby}}\ and\ \bibinfo {author} {\bibfnamefont {J.}~\bibnamefont
  {Klassen}},\ }\bibfield  {title} {\bibinfo {title} {A compact fermion to
  qubit mapping part 2: Alternative lattice geometries},\ }\href@noop {}
  {\bibfield  {journal} {\bibinfo  {journal} {arXiv:2101.10735}\ } (\bibinfo
  {year} {2021})}\BibitemShut {NoStop}%
\bibitem [{\citenamefont {Landahl}\ and\ \citenamefont
  {Morrison}(2021)}]{landahl2021logical}%
  \BibitemOpen
  \bibfield  {author} {\bibinfo {author} {\bibfnamefont {A.~J.}\ \bibnamefont
  {Landahl}}\ and\ \bibinfo {author} {\bibfnamefont {B.~C.}\ \bibnamefont
  {Morrison}},\ }\bibfield  {title} {\bibinfo {title} {Logical majorana
  fermions for fault-tolerant quantum simulation},\ }\href@noop {} {\bibfield
  {journal} {\bibinfo  {journal} {arXiv:2110.10280}\ } (\bibinfo {year}
  {2021})}\BibitemShut {NoStop}%
\bibitem [{\citenamefont {Chen}\ and\ \citenamefont
  {Xu}(2022)}]{chen2022equivalence}%
  \BibitemOpen
  \bibfield  {author} {\bibinfo {author} {\bibfnamefont {Y.-A.}\ \bibnamefont
  {Chen}}\ and\ \bibinfo {author} {\bibfnamefont {Y.}~\bibnamefont {Xu}},\
  }\bibfield  {title} {\bibinfo {title} {Equivalence between fermion-to-qubit
  mappings in two spatial dimensions},\ }\href@noop {} {\bibfield  {journal}
  {\bibinfo  {journal} {arXiv:2201.05153}\ } (\bibinfo {year}
  {2022})}\BibitemShut {NoStop}%
\bibitem [{\citenamefont {Chien}\ and\ \citenamefont
  {Klassen}(2022)}]{chien2022optimizing}%
  \BibitemOpen
  \bibfield  {author} {\bibinfo {author} {\bibfnamefont {R.~W.}\ \bibnamefont
  {Chien}}\ and\ \bibinfo {author} {\bibfnamefont {J.}~\bibnamefont
  {Klassen}},\ }\bibfield  {title} {\bibinfo {title} {Optimizing fermionic
  encodings for both hamiltonian and hardware},\ }\href@noop {} {\bibfield
  {journal} {\bibinfo  {journal} {arXiv:2210.05652}\ } (\bibinfo {year}
  {2022})}\BibitemShut {NoStop}%
\bibitem [{\citenamefont {Chen}\ \emph {et~al.}(2022)\citenamefont {Chen},
  \citenamefont {Gorshkov},\ and\ \citenamefont {Xu}}]{chen2022error}%
  \BibitemOpen
  \bibfield  {author} {\bibinfo {author} {\bibfnamefont {Y.-A.}\ \bibnamefont
  {Chen}}, \bibinfo {author} {\bibfnamefont {A.~V.}\ \bibnamefont {Gorshkov}},\
  and\ \bibinfo {author} {\bibfnamefont {Y.}~\bibnamefont {Xu}},\ }\bibfield
  {title} {\bibinfo {title} {Error-correcting codes for fermionic quantum
  simulation},\ }\href@noop {} {\bibfield  {journal} {\bibinfo  {journal}
  {arXiv:2210.08411}\ } (\bibinfo {year} {2022})}\BibitemShut {NoStop}%
\bibitem [{\citenamefont {Hu}\ \emph {et~al.}(2019)\citenamefont {Hu},
  \citenamefont {Ma}, \citenamefont {Cai}, \citenamefont {Mu}, \citenamefont
  {Xu}, \citenamefont {Wang}, \citenamefont {Wu}, \citenamefont {Wang},
  \citenamefont {Song}, \citenamefont {Zou} \emph {et~al.}}]{hu2019quantum}%
  \BibitemOpen
  \bibfield  {author} {\bibinfo {author} {\bibfnamefont {L.}~\bibnamefont
  {Hu}}, \bibinfo {author} {\bibfnamefont {Y.}~\bibnamefont {Ma}}, \bibinfo
  {author} {\bibfnamefont {W.}~\bibnamefont {Cai}}, \bibinfo {author}
  {\bibfnamefont {X.}~\bibnamefont {Mu}}, \bibinfo {author} {\bibfnamefont
  {Y.}~\bibnamefont {Xu}}, \bibinfo {author} {\bibfnamefont {W.}~\bibnamefont
  {Wang}}, \bibinfo {author} {\bibfnamefont {Y.}~\bibnamefont {Wu}}, \bibinfo
  {author} {\bibfnamefont {H.}~\bibnamefont {Wang}}, \bibinfo {author}
  {\bibfnamefont {Y.}~\bibnamefont {Song}}, \bibinfo {author} {\bibfnamefont
  {C.-L.}\ \bibnamefont {Zou}}, \emph {et~al.},\ }\bibfield  {title} {\bibinfo
  {title} {Quantum error correction and universal gate set operation on a
  binomial bosonic logical qubit},\ }\href@noop {} {\bibfield  {journal}
  {\bibinfo  {journal} {Nature Physics}\ }\textbf {\bibinfo {volume} {15}},\
  \bibinfo {pages} {503} (\bibinfo {year} {2019})}\BibitemShut {NoStop}%
\bibitem [{\citenamefont {Andersen}\ \emph {et~al.}(2020)\citenamefont
  {Andersen}, \citenamefont {Remm}, \citenamefont {Lazar}, \citenamefont
  {Krinner}, \citenamefont {Lacroix}, \citenamefont {Norris}, \citenamefont
  {Gabureac}, \citenamefont {Eichler},\ and\ \citenamefont
  {Wallraff}}]{andersen2020repeated}%
  \BibitemOpen
  \bibfield  {author} {\bibinfo {author} {\bibfnamefont {C.~K.}\ \bibnamefont
  {Andersen}}, \bibinfo {author} {\bibfnamefont {A.}~\bibnamefont {Remm}},
  \bibinfo {author} {\bibfnamefont {S.}~\bibnamefont {Lazar}}, \bibinfo
  {author} {\bibfnamefont {S.}~\bibnamefont {Krinner}}, \bibinfo {author}
  {\bibfnamefont {N.}~\bibnamefont {Lacroix}}, \bibinfo {author} {\bibfnamefont
  {G.~J.}\ \bibnamefont {Norris}}, \bibinfo {author} {\bibfnamefont
  {M.}~\bibnamefont {Gabureac}}, \bibinfo {author} {\bibfnamefont
  {C.}~\bibnamefont {Eichler}},\ and\ \bibinfo {author} {\bibfnamefont
  {A.}~\bibnamefont {Wallraff}},\ }\bibfield  {title} {\bibinfo {title}
  {Repeated quantum error detection in a surface code},\ }\href@noop {}
  {\bibfield  {journal} {\bibinfo  {journal} {Nature Physics}\ }\textbf
  {\bibinfo {volume} {16}},\ \bibinfo {pages} {875} (\bibinfo {year}
  {2020})}\BibitemShut {NoStop}%
\bibitem [{\citenamefont {Egan}\ \emph {et~al.}(2021)\citenamefont {Egan},
  \citenamefont {Debroy}, \citenamefont {Noel}, \citenamefont {Risinger},
  \citenamefont {Zhu}, \citenamefont {Biswas}, \citenamefont {Newman},
  \citenamefont {Li}, \citenamefont {Brown}, \citenamefont {Cetina} \emph
  {et~al.}}]{egan2021fault}%
  \BibitemOpen
  \bibfield  {author} {\bibinfo {author} {\bibfnamefont {L.}~\bibnamefont
  {Egan}}, \bibinfo {author} {\bibfnamefont {D.~M.}\ \bibnamefont {Debroy}},
  \bibinfo {author} {\bibfnamefont {C.}~\bibnamefont {Noel}}, \bibinfo {author}
  {\bibfnamefont {A.}~\bibnamefont {Risinger}}, \bibinfo {author}
  {\bibfnamefont {D.}~\bibnamefont {Zhu}}, \bibinfo {author} {\bibfnamefont
  {D.}~\bibnamefont {Biswas}}, \bibinfo {author} {\bibfnamefont
  {M.}~\bibnamefont {Newman}}, \bibinfo {author} {\bibfnamefont
  {M.}~\bibnamefont {Li}}, \bibinfo {author} {\bibfnamefont {K.~R.}\
  \bibnamefont {Brown}}, \bibinfo {author} {\bibfnamefont {M.}~\bibnamefont
  {Cetina}}, \emph {et~al.},\ }\bibfield  {title} {\bibinfo {title}
  {Fault-tolerant control of an error-corrected qubit},\ }\href@noop {}
  {\bibfield  {journal} {\bibinfo  {journal} {Nature}\ }\textbf {\bibinfo
  {volume} {598}},\ \bibinfo {pages} {281} (\bibinfo {year}
  {2021})}\BibitemShut {NoStop}%
\bibitem [{\citenamefont {Ryan-Anderson}\ \emph {et~al.}(2021)\citenamefont
  {Ryan-Anderson}, \citenamefont {Bohnet}, \citenamefont {Lee}, \citenamefont
  {Gresh}, \citenamefont {Hankin}, \citenamefont {Gaebler}, \citenamefont
  {Francois}, \citenamefont {Chernoguzov}, \citenamefont {Lucchetti},
  \citenamefont {Brown}, \citenamefont {Gatterman}, \citenamefont {Halit},
  \citenamefont {Gilmore}, \citenamefont {Gerber}, \citenamefont {Neyenhuis},
  \citenamefont {Hayes},\ and\ \citenamefont {Stutz}}]{QuantinuumEC}%
  \BibitemOpen
  \bibfield  {author} {\bibinfo {author} {\bibfnamefont {C.}~\bibnamefont
  {Ryan-Anderson}}, \bibinfo {author} {\bibfnamefont {J.~G.}\ \bibnamefont
  {Bohnet}}, \bibinfo {author} {\bibfnamefont {K.}~\bibnamefont {Lee}},
  \bibinfo {author} {\bibfnamefont {D.}~\bibnamefont {Gresh}}, \bibinfo
  {author} {\bibfnamefont {A.}~\bibnamefont {Hankin}}, \bibinfo {author}
  {\bibfnamefont {J.~P.}\ \bibnamefont {Gaebler}}, \bibinfo {author}
  {\bibfnamefont {D.}~\bibnamefont {Francois}}, \bibinfo {author}
  {\bibfnamefont {A.}~\bibnamefont {Chernoguzov}}, \bibinfo {author}
  {\bibfnamefont {D.}~\bibnamefont {Lucchetti}}, \bibinfo {author}
  {\bibfnamefont {N.~C.}\ \bibnamefont {Brown}}, \bibinfo {author}
  {\bibfnamefont {T.~M.}\ \bibnamefont {Gatterman}}, \bibinfo {author}
  {\bibfnamefont {S.~K.}\ \bibnamefont {Halit}}, \bibinfo {author}
  {\bibfnamefont {K.}~\bibnamefont {Gilmore}}, \bibinfo {author} {\bibfnamefont
  {J.~A.}\ \bibnamefont {Gerber}}, \bibinfo {author} {\bibfnamefont
  {B.}~\bibnamefont {Neyenhuis}}, \bibinfo {author} {\bibfnamefont
  {D.}~\bibnamefont {Hayes}},\ and\ \bibinfo {author} {\bibfnamefont {R.~P.}\
  \bibnamefont {Stutz}},\ }\bibfield  {title} {\bibinfo {title} {Realization of
  real-time fault-tolerant quantum error correction},\ }\href
  {https://doi.org/10.1103/PhysRevX.11.041058} {\bibfield  {journal} {\bibinfo
  {journal} {Phys. Rev. X}\ }\textbf {\bibinfo {volume} {11}},\ \bibinfo
  {pages} {041058} (\bibinfo {year} {2021})}\BibitemShut {NoStop}%
\bibitem [{\citenamefont {Acharya}\ \emph {et~al.}(2023)\citenamefont
  {Acharya}, \citenamefont {Aleiner}, \citenamefont {Allen}, \citenamefont
  {Andersen}, \citenamefont {Ansmann}, \citenamefont {Arute}, \citenamefont
  {Arya}, \citenamefont {Asfaw}, \citenamefont {Atalaya}, \citenamefont
  {Babbush} \emph {et~al.}}]{google2023suppressing}%
  \BibitemOpen
  \bibfield  {author} {\bibinfo {author} {\bibfnamefont {R.}~\bibnamefont
  {Acharya}}, \bibinfo {author} {\bibfnamefont {I.}~\bibnamefont {Aleiner}},
  \bibinfo {author} {\bibfnamefont {R.}~\bibnamefont {Allen}}, \bibinfo
  {author} {\bibfnamefont {T.~I.}\ \bibnamefont {Andersen}}, \bibinfo {author}
  {\bibfnamefont {M.}~\bibnamefont {Ansmann}}, \bibinfo {author} {\bibfnamefont
  {F.}~\bibnamefont {Arute}}, \bibinfo {author} {\bibfnamefont
  {K.}~\bibnamefont {Arya}}, \bibinfo {author} {\bibfnamefont {A.}~\bibnamefont
  {Asfaw}}, \bibinfo {author} {\bibfnamefont {J.}~\bibnamefont {Atalaya}},
  \bibinfo {author} {\bibfnamefont {R.}~\bibnamefont {Babbush}}, \emph
  {et~al.},\ }\bibfield  {title} {\bibinfo {title} {Suppressing quantum errors
  by scaling a surface code logical qubit},\ }\href@noop {} {\bibfield
  {journal} {\bibinfo  {journal} {Nature}\ }\textbf {\bibinfo {volume} {614}},\
  \bibinfo {pages} {676} (\bibinfo {year} {2023})}\BibitemShut {NoStop}%
\bibitem [{\citenamefont {Viola}\ \emph {et~al.}(1999)\citenamefont {Viola},
  \citenamefont {Knill},\ and\ \citenamefont {Lloyd}}]{viola1999dynamical}%
  \BibitemOpen
  \bibfield  {author} {\bibinfo {author} {\bibfnamefont {L.}~\bibnamefont
  {Viola}}, \bibinfo {author} {\bibfnamefont {E.}~\bibnamefont {Knill}},\ and\
  \bibinfo {author} {\bibfnamefont {S.}~\bibnamefont {Lloyd}},\ }\bibfield
  {title} {\bibinfo {title} {Dynamical decoupling of open quantum systems},\
  }\href@noop {} {\bibfield  {journal} {\bibinfo  {journal} {Physical Review
  Letters}\ }\textbf {\bibinfo {volume} {82}},\ \bibinfo {pages} {2417}
  (\bibinfo {year} {1999})}\BibitemShut {NoStop}%
\bibitem [{\citenamefont {Temme}\ \emph {et~al.}(2017)\citenamefont {Temme},
  \citenamefont {Bravyi},\ and\ \citenamefont {Gambetta}}]{temme2017error}%
  \BibitemOpen
  \bibfield  {author} {\bibinfo {author} {\bibfnamefont {K.}~\bibnamefont
  {Temme}}, \bibinfo {author} {\bibfnamefont {S.}~\bibnamefont {Bravyi}},\ and\
  \bibinfo {author} {\bibfnamefont {J.~M.}\ \bibnamefont {Gambetta}},\
  }\bibfield  {title} {\bibinfo {title} {Error mitigation for short-depth
  quantum circuits},\ }\href@noop {} {\bibfield  {journal} {\bibinfo  {journal}
  {Physical review letters}\ }\textbf {\bibinfo {volume} {119}},\ \bibinfo
  {pages} {180509} (\bibinfo {year} {2017})}\BibitemShut {NoStop}%
\bibitem [{\citenamefont {Lowe}\ \emph {et~al.}(2021)\citenamefont {Lowe},
  \citenamefont {Gordon}, \citenamefont {Czarnik}, \citenamefont {Arrasmith},
  \citenamefont {Coles},\ and\ \citenamefont {Cincio}}]{lowe2021unified}%
  \BibitemOpen
  \bibfield  {author} {\bibinfo {author} {\bibfnamefont {A.}~\bibnamefont
  {Lowe}}, \bibinfo {author} {\bibfnamefont {M.~H.}\ \bibnamefont {Gordon}},
  \bibinfo {author} {\bibfnamefont {P.}~\bibnamefont {Czarnik}}, \bibinfo
  {author} {\bibfnamefont {A.}~\bibnamefont {Arrasmith}}, \bibinfo {author}
  {\bibfnamefont {P.~J.}\ \bibnamefont {Coles}},\ and\ \bibinfo {author}
  {\bibfnamefont {L.}~\bibnamefont {Cincio}},\ }\bibfield  {title} {\bibinfo
  {title} {Unified approach to data-driven quantum error mitigation},\
  }\href@noop {} {\bibfield  {journal} {\bibinfo  {journal} {Physical Review
  Research}\ }\textbf {\bibinfo {volume} {3}},\ \bibinfo {pages} {033098}
  (\bibinfo {year} {2021})}\BibitemShut {NoStop}%
\bibitem [{\citenamefont {Huggins}\ \emph {et~al.}(2021)\citenamefont
  {Huggins}, \citenamefont {McArdle}, \citenamefont {O’Brien}, \citenamefont
  {Lee}, \citenamefont {Rubin}, \citenamefont {Boixo}, \citenamefont {Whaley},
  \citenamefont {Babbush},\ and\ \citenamefont {McClean}}]{huggins2021virtual}%
  \BibitemOpen
  \bibfield  {author} {\bibinfo {author} {\bibfnamefont {W.~J.}\ \bibnamefont
  {Huggins}}, \bibinfo {author} {\bibfnamefont {S.}~\bibnamefont {McArdle}},
  \bibinfo {author} {\bibfnamefont {T.~E.}\ \bibnamefont {O’Brien}}, \bibinfo
  {author} {\bibfnamefont {J.}~\bibnamefont {Lee}}, \bibinfo {author}
  {\bibfnamefont {N.~C.}\ \bibnamefont {Rubin}}, \bibinfo {author}
  {\bibfnamefont {S.}~\bibnamefont {Boixo}}, \bibinfo {author} {\bibfnamefont
  {K.~B.}\ \bibnamefont {Whaley}}, \bibinfo {author} {\bibfnamefont
  {R.}~\bibnamefont {Babbush}},\ and\ \bibinfo {author} {\bibfnamefont {J.~R.}\
  \bibnamefont {McClean}},\ }\bibfield  {title} {\bibinfo {title} {Virtual
  distillation for quantum error mitigation},\ }\href@noop {} {\bibfield
  {journal} {\bibinfo  {journal} {Physical Review X}\ }\textbf {\bibinfo
  {volume} {11}},\ \bibinfo {pages} {041036} (\bibinfo {year}
  {2021})}\BibitemShut {NoStop}%
\bibitem [{\citenamefont {Bonet-Monroig}\ \emph {et~al.}(2018)\citenamefont
  {Bonet-Monroig}, \citenamefont {Sagastizabal}, \citenamefont {Singh},\ and\
  \citenamefont {O'Brien}}]{bonet2018low}%
  \BibitemOpen
  \bibfield  {author} {\bibinfo {author} {\bibfnamefont {X.}~\bibnamefont
  {Bonet-Monroig}}, \bibinfo {author} {\bibfnamefont {R.}~\bibnamefont
  {Sagastizabal}}, \bibinfo {author} {\bibfnamefont {M.}~\bibnamefont
  {Singh}},\ and\ \bibinfo {author} {\bibfnamefont {T.}~\bibnamefont
  {O'Brien}},\ }\bibfield  {title} {\bibinfo {title} {Low-cost error mitigation
  by symmetry verification},\ }\href@noop {} {\bibfield  {journal} {\bibinfo
  {journal} {Physical Review A}\ }\textbf {\bibinfo {volume} {98}},\ \bibinfo
  {pages} {062339} (\bibinfo {year} {2018})}\BibitemShut {NoStop}%
\bibitem [{\citenamefont {McClean}\ \emph {et~al.}(2020)\citenamefont
  {McClean}, \citenamefont {Jiang}, \citenamefont {Rubin}, \citenamefont
  {Babbush},\ and\ \citenamefont {Neven}}]{mcclean2020decoding}%
  \BibitemOpen
  \bibfield  {author} {\bibinfo {author} {\bibfnamefont {J.~R.}\ \bibnamefont
  {McClean}}, \bibinfo {author} {\bibfnamefont {Z.}~\bibnamefont {Jiang}},
  \bibinfo {author} {\bibfnamefont {N.~C.}\ \bibnamefont {Rubin}}, \bibinfo
  {author} {\bibfnamefont {R.}~\bibnamefont {Babbush}},\ and\ \bibinfo {author}
  {\bibfnamefont {H.}~\bibnamefont {Neven}},\ }\bibfield  {title} {\bibinfo
  {title} {Decoding quantum errors with subspace expansions},\ }\href@noop {}
  {\bibfield  {journal} {\bibinfo  {journal} {Nature communications}\ }\textbf
  {\bibinfo {volume} {11}},\ \bibinfo {pages} {1} (\bibinfo {year}
  {2020})}\BibitemShut {NoStop}%
\bibitem [{\citenamefont {McArdle}\ \emph {et~al.}(2019)\citenamefont
  {McArdle}, \citenamefont {Yuan},\ and\ \citenamefont
  {Benjamin}}]{mcardle2019error}%
  \BibitemOpen
  \bibfield  {author} {\bibinfo {author} {\bibfnamefont {S.}~\bibnamefont
  {McArdle}}, \bibinfo {author} {\bibfnamefont {X.}~\bibnamefont {Yuan}},\ and\
  \bibinfo {author} {\bibfnamefont {S.}~\bibnamefont {Benjamin}},\ }\bibfield
  {title} {\bibinfo {title} {Error-mitigated digital quantum simulation},\
  }\href@noop {} {\bibfield  {journal} {\bibinfo  {journal} {Physical review
  letters}\ }\textbf {\bibinfo {volume} {122}},\ \bibinfo {pages} {180501}
  (\bibinfo {year} {2019})}\BibitemShut {NoStop}%
\bibitem [{\citenamefont {Chen}(2020)}]{chen2020arbitrary}%
  \BibitemOpen
  \bibfield  {author} {\bibinfo {author} {\bibfnamefont {Y.-A.}\ \bibnamefont
  {Chen}},\ }\bibfield  {title} {\bibinfo {title} {Exact bosonization in
  arbitrary dimensions},\ }\href@noop {} {\bibfield  {journal} {\bibinfo
  {journal} {Physical Review Research}\ }\textbf {\bibinfo {volume} {2}},\
  \bibinfo {pages} {033527} (\bibinfo {year} {2020})}\BibitemShut {NoStop}%
\bibitem [{\citenamefont {Wigner}\ and\ \citenamefont
  {Jordan}(1928)}]{wigner1928paulische}%
  \BibitemOpen
  \bibfield  {author} {\bibinfo {author} {\bibfnamefont {E.}~\bibnamefont
  {Wigner}}\ and\ \bibinfo {author} {\bibfnamefont {P.}~\bibnamefont
  {Jordan}},\ }\bibfield  {title} {\bibinfo {title} {{\"U}ber das paulische
  {\"a}quivalenzverbot},\ }\href@noop {} {\bibfield  {journal} {\bibinfo
  {journal} {Z. Phys}\ }\textbf {\bibinfo {volume} {47}},\ \bibinfo {pages}
  {631} (\bibinfo {year} {1928})}\BibitemShut {NoStop}%
\bibitem [{\citenamefont {Bravyi}\ \emph {et~al.}(2017)\citenamefont {Bravyi},
  \citenamefont {Gambetta}, \citenamefont {Mezzacapo},\ and\ \citenamefont
  {Temme}}]{bravyi2017tapering}%
  \BibitemOpen
  \bibfield  {author} {\bibinfo {author} {\bibfnamefont {S.}~\bibnamefont
  {Bravyi}}, \bibinfo {author} {\bibfnamefont {J.~M.}\ \bibnamefont
  {Gambetta}}, \bibinfo {author} {\bibfnamefont {A.}~\bibnamefont
  {Mezzacapo}},\ and\ \bibinfo {author} {\bibfnamefont {K.}~\bibnamefont
  {Temme}},\ }\bibfield  {title} {\bibinfo {title} {Tapering off qubits to
  simulate fermionic hamiltonians},\ }\href@noop {} {\bibfield  {journal}
  {\bibinfo  {journal} {arXiv:1701.08213}\ } (\bibinfo {year}
  {2017})}\BibitemShut {NoStop}%
\bibitem [{\citenamefont {Setia}\ \emph {et~al.}(2020)\citenamefont {Setia},
  \citenamefont {Chen}, \citenamefont {Rice}, \citenamefont {Mezzacapo},
  \citenamefont {Pistoia},\ and\ \citenamefont
  {Whitfield}}]{setia2020reducing}%
  \BibitemOpen
  \bibfield  {author} {\bibinfo {author} {\bibfnamefont {K.}~\bibnamefont
  {Setia}}, \bibinfo {author} {\bibfnamefont {R.}~\bibnamefont {Chen}},
  \bibinfo {author} {\bibfnamefont {J.~E.}\ \bibnamefont {Rice}}, \bibinfo
  {author} {\bibfnamefont {A.}~\bibnamefont {Mezzacapo}}, \bibinfo {author}
  {\bibfnamefont {M.}~\bibnamefont {Pistoia}},\ and\ \bibinfo {author}
  {\bibfnamefont {J.~D.}\ \bibnamefont {Whitfield}},\ }\bibfield  {title}
  {\bibinfo {title} {Reducing qubit requirements for quantum simulations using
  molecular point group symmetries},\ }\href@noop {} {\bibfield  {journal}
  {\bibinfo  {journal} {Journal of Chemical Theory and Computation}\ }\textbf
  {\bibinfo {volume} {16}},\ \bibinfo {pages} {6091} (\bibinfo {year}
  {2020})}\BibitemShut {NoStop}%
\bibitem [{\citenamefont {Wierichs}\ \emph {et~al.}(2022)\citenamefont
  {Wierichs}, \citenamefont {Izaac}, \citenamefont {Wang},\ and\ \citenamefont
  {Lin}}]{wierichs2022general}%
  \BibitemOpen
  \bibfield  {author} {\bibinfo {author} {\bibfnamefont {D.}~\bibnamefont
  {Wierichs}}, \bibinfo {author} {\bibfnamefont {J.}~\bibnamefont {Izaac}},
  \bibinfo {author} {\bibfnamefont {C.}~\bibnamefont {Wang}},\ and\ \bibinfo
  {author} {\bibfnamefont {C.~Y.-Y.}\ \bibnamefont {Lin}},\ }\bibfield  {title}
  {\bibinfo {title} {General parameter-shift rules for quantum gradients},\
  }\href@noop {} {\bibfield  {journal} {\bibinfo  {journal} {Quantum}\ }\textbf
  {\bibinfo {volume} {6}},\ \bibinfo {pages} {677} (\bibinfo {year}
  {2022})}\BibitemShut {NoStop}%
\bibitem [{\citenamefont {Rudner}\ \emph {et~al.}(2013)\citenamefont {Rudner},
  \citenamefont {Lindner}, \citenamefont {Berg},\ and\ \citenamefont
  {Levin}}]{rudner2013anomalous}%
  \BibitemOpen
  \bibfield  {author} {\bibinfo {author} {\bibfnamefont {M.~S.}\ \bibnamefont
  {Rudner}}, \bibinfo {author} {\bibfnamefont {N.~H.}\ \bibnamefont {Lindner}},
  \bibinfo {author} {\bibfnamefont {E.}~\bibnamefont {Berg}},\ and\ \bibinfo
  {author} {\bibfnamefont {M.}~\bibnamefont {Levin}},\ }\bibfield  {title}
  {\bibinfo {title} {Anomalous edge states and the bulk-edge correspondence for
  periodically driven two-dimensional systems},\ }\href@noop {} {\bibfield
  {journal} {\bibinfo  {journal} {Physical Review X}\ }\textbf {\bibinfo
  {volume} {3}},\ \bibinfo {pages} {031005} (\bibinfo {year}
  {2013})}\BibitemShut {NoStop}%
\bibitem [{\citenamefont {Chien}\ \emph {et~al.}(2019)\citenamefont {Chien},
  \citenamefont {Xue}, \citenamefont {Hardikar}, \citenamefont {Setia},\ and\
  \citenamefont {Whitfield}}]{chien2019analysis}%
  \BibitemOpen
  \bibfield  {author} {\bibinfo {author} {\bibfnamefont {R.~W.}\ \bibnamefont
  {Chien}}, \bibinfo {author} {\bibfnamefont {S.}~\bibnamefont {Xue}}, \bibinfo
  {author} {\bibfnamefont {T.~S.}\ \bibnamefont {Hardikar}}, \bibinfo {author}
  {\bibfnamefont {K.}~\bibnamefont {Setia}},\ and\ \bibinfo {author}
  {\bibfnamefont {J.~D.}\ \bibnamefont {Whitfield}},\ }\bibfield  {title}
  {\bibinfo {title} {Analysis of superfast encoding performance for electronic
  structure simulations},\ }\href@noop {} {\bibfield  {journal} {\bibinfo
  {journal} {Physical Review A}\ }\textbf {\bibinfo {volume} {100}},\ \bibinfo
  {pages} {032337} (\bibinfo {year} {2019})}\BibitemShut {NoStop}%
\end{thebibliography}%
\end{document}